\documentclass[conference]{IEEEtran}
\IEEEoverridecommandlockouts
\usepackage{cite}
\usepackage{amsmath,amssymb,amsfonts}
\usepackage{algorithmic}

\usepackage{graphicx}
\usepackage{textcomp}
\usepackage{xcolor}
\usepackage{makecell}
\usepackage{ulem}
\usepackage{booktabs}
\usepackage{threeparttable}
\usepackage{hyperref} 
\usepackage{algorithm}
\usepackage{multirow}
\def\BibTeX{{\rm B\kern-.05em{\sc i\kern-.025em b}\kern-.08em
    T\kern-.1667em\lower.7ex\hbox{E}\kern-.125emX}}
\begin{document}

\title{ENSI: Efficient Non-Interactive Secure Inference for Large Language Models
\thanks{Yuchuan Luo is the corresponding author.}
}
% \author{\IEEEauthorblockN{Anonymous Authors}
% \IEEEauthorblockA{\textit{Affiliation}}
% \IEEEauthorblockA{\textit{Email}}
% }

\author{\IEEEauthorblockN{Zhiyu He, Maojiang Wang, Xinwen Gao, Yuchuan Luo*, Lin Liu and Shaojing Fu}
\IEEEauthorblockA{\textit{College of Computer Science and Technology} \\
\textit{National University of Defense Technology}\\
Changsha, Hunan, China\\
\{hezhiyu99, wangmaojiang19, gaoxinwen17, luoyuchuan09, liulin16, fushaojing\}@nudt.edu.cn}}

% \and
% \IEEEauthorblockN{2\textsuperscript{nd} Given Name Surname}
% \IEEEauthorblockA{\textit{dept. name of organization (of Aff.)} \\
% \textit{name of organization (of Aff.)}\\
% City, Country \\
% email address or ORCID}
% \and
% \IEEEauthorblockN{3\textsuperscript{rd} Given Name Surname}
% \IEEEauthorblockA{\textit{dept. name of organization (of Aff.)} \\
% \textit{name of organization (of Aff.)}\\
% City, Country \\
% email address or ORCID}
% \and
% \IEEEauthorblockN{4\textsuperscript{th} Given Name Surname}
% \IEEEauthorblockA{\textit{dept. name of organization (of Aff.)} \\
% \textit{name of organization (of Aff.)}\\
% City, Country \\
% email address or ORCID}
% \and
% \IEEEauthorblockN{5\textsuperscript{th} Given Name Surname}
% \IEEEauthorblockA{\textit{dept. name of organization (of Aff.)} \\
% \textit{name of organization (of Aff.)}\\
% City, Country \\
% email address or ORCID}
% \and
% \IEEEauthorblockN{6\textsuperscript{th} Given Name Surname}
% \IEEEauthorblockA{\textit{dept. name of organization (of Aff.)} \\
% \textit{name of organization (of Aff.)}\\
% City, Country \\
% email address or ORCID}

\maketitle

\begin{abstract}
Secure inference enables privacy-preserving machine learning by leveraging cryptographic protocols that support computations on sensitive user data without exposing it. However, integrating cryptographic protocols with large language models (LLMs) presents significant challenges, as the inherent complexity of these protocols, together with LLMs' massive parameter scale and sophisticated architectures, severely limits practical usability.
% However, at the convergence of cryptographic protocols and large language models (LLMs), the intrinsic complexity of the protocols combined with LLMs' massive parameter scale and intricate architectures severely limits practical usability.

% Secure inference enables privacy-preserving deployment of machine learning models by allowing computations on sensitive user data without exposing it. However, cryptographic protocols introduces complex computational primitives, while the large-scale parameters and complex architecture of large language models further constrain the feasibility and efficiency.

In this work, we propose ENSI, a novel non-interactive secure inference framework for LLMs, based on the principle of co-designing the cryptographic protocols and LLM architecture. ENSI employs an optimized encoding strategy that seamlessly integrates CKKS scheme with a lightweight LLM variant, BitNet, significantly reducing the computational complexity of encrypted matrix multiplications. 
% Given the prohibitive cost of the traditional softmax operation under homomorphic encryption, we introduce the sigmoid-attention mechanism as drop-in replacement, which requires no additional model retraining. 
In response to the prohibitive computational demands of softmax under homomorphic encryption (HE), we pioneer the integration of the sigmoid attention mechanism with HE as a seamless, retraining-free alternative. Furthermore, by embedding the Bootstrapping operation within the RMSNorm process, we efficiently refresh ciphertexts while markedly decreasing the frequency of costly bootstrapping invocations. 
Experimental evaluations demonstrate that ENSI achieves approximately an $8\times$ acceleration in matrix multiplications and a $2.6\times$ speedup in softmax inference on CPU compared to state-of-the-art method, with the proportion of bootstrapping is reduced to just $1\%$.

\end{abstract}


\section{Introduction}
% %诸如LLaMA这样的Transformer模型开启了人工智能领域的新时代。这些被广泛采用的大型语言模型（LLM）最显著的能力之一，是根据用户的独特需求提供量身定制的个性化回复。然而，这种显著的便利性不可避免地涉及对敏感数据的处理，从而引发了一系列安全问题。例如，随着LLM在推理过程中越来越多地处理用户特定的信息，缺乏有效的安全机制可能导致敏感信息在无意中暴露给不应接收的对象。
% Transformer-based models, such as LLaMA\cite{Touvron2023} and GPT\cite{Radford2019}, have inaugurated a new era in the field of artificial intelligence. Among the most prominent capabilities of these widely adopted large language models (LLMs) is their ability to deliver personalized responses tailored to the unique needs of users. However, this remarkable convenience inherently involves the handling of sensitive data, giving rise to a variety of security concerns. For instance, as LLMs increasingly process user-specific information during inference, the absence of robust security mechanisms can result in the inadvertent exposure of sensitive details to unintended recipients\cite{Mire2023}.

%基于变压器的模型，如LLaMA\cite{Touvron2023}和GPT\cite{Radford2019}，开创了人工智能领域的新时代。这些被广泛采用的大型语言模型（LLM）最突出的功能之一是能够根据用户的独特需求提供个性化的响应。这种能力催生了新兴的“推理即服务”（IaaS）范式。通过IaaS，用户可以免除自己部署复杂和资源密集型模型的负担，而是通过基于云的API实时访问这些强大的语言模型来完成各种任务。然而，随着LLM在推理过程中越来越多地处理用户特定的敏感数据，缺乏强大的安全机制可能会导致此类敏感信息无意中泄露给未经授权的各方{Mire2023}。
Large Language Models, such as LLaMA\cite{Touvron2023, touvron2023llama2} and GPT\cite{GPT}, have inaugurated a new era in the field of artificial intelligence. Among the most prominent capabilities of these widely adopted large language models is their ability to deliver personalized responses tailored to the unique needs of users.
This capability gives rise to the emerging paradigm of Inference as a Service (IaaS). Through IaaS, users are relieved from the burden of deploying complex and resource-intensive models themselves, instead accessing these powerful language models in real time via cloud-based APIs to accomplish diverse tasks. However, as LLMs increasingly process user-specific sensitive data during inference, the absence of robust security mechanisms may result in inadvertent disclosure of such sensitive information to unauthorized parties\cite{Mire2023}.

%安全推理作为一种新兴的计算范式，正成为推进隐私保护机器学习的关键驱动力。客户端（用户）对其私有数据进行加密，而服务器通过专门的协议对密文进行计算，而无需访问底层明文。为了实现安全推理，主要采用了两种具有可证明安全性的主要加密技术：安全多方计算（MPC）和同态加密（HE）。MPC需要多轮通信，最适合具有高网络稳定性的环境。相比之下，HE利用其完全同态特性，实现了非交互式计算外包，提供了更大的通用适应性和灵活性。
Secure inference, as an emerging computational paradigm, is becoming a key driving force for advancing privacy-preserving machine learning. Clients (users) encrypt their private data, while servers perform computations on the ciphertexts through specialized protocols without accessing the users’ secret plaintext data.
To enable secure inference, two main cryptographic techniques with provable security are predominantly employed: Secure Multi-Party Computation (SMPC)\cite{Yao1982} and Homomorphic Encryption\cite{Gentry2009}. SMPC, requiring multiple rounds of communication, is best suited for environments with high network stability. In contrast, HE, leveraging its fully homomorphic properties, enables non-interactive computation outsourcing, offering greater general adaptability and flexibility. In particular, non-interactive secure inference based on HE holds significant promise for distributed environments, exhibiting marked advantages in both computational efficiency and privacy preservation.

%尽管在CNN等传统神经网络的安全推理方面取得了重大进展，但应用HE在Transformer模型中实现安全推理仍然极具挑战性。一方面，HE背后的复杂计算原语大大增加了密文算术运算的计算开销，密文算术运算总是比明文运算慢得多。另一方面，由于其参数规模大和复杂的网络架构，Transformer模型需要比CNN高得多的计算和存储资源。特别值得注意的是变压器中非线性激活函数的广泛使用，众所周知，变压器对HE环境不友好。因此，实现LLM的高效安全推理迫切需要一种包括模型架构设计和密码算法开发的联合优化方法。

% Despite significant advances in secure inference for traditional neuralnetworks like CNNs\cite{Juvekar2018, AlBadawi2021, Huang2022}, applying HE to enable secure inference in Transformer models remains highly challenging. (1) The complex computational primitives underlying HE substantially increase the computational overhead of ciphertext arithmetic operations, which are invariably much slower than their plaintext counterparts. (2) Transformer models demand considerably higher computational and storage resources than CNNs due to their large parameter scale and intricate network architectures. Particularly noteworthy is the widespread use of nonlinear activation functions in Transformers, which are notoriously unfriendly to HE environments. Consequently, achieving efficient secure inference for LLMs urgently calls for a joint optimization approach encompassing both model architecture design and cryptographic algorithm development.

Despite significant advances in secure inference for traditional neural networks like CNNs\cite{Juvekar2018, AlBadawi2021, Huang2022}, applying HE to enable secure inference in large language models remains highly challenging.
\begin{itemize}
\item[1)]LLMs rely heavily on high-dimensional matrix multiplications and stacked self-attention mechanisms, demanding substantially more computational resources than CNNs. In addition, the sophisticated activation functions commonly used in LLMs are particularly problematic, as they are notoriously difficult to implement efficiently in HE environments.
%Particularly noteworthy is the use of activation functions in LLMs, which are notoriously unfriendly to HE environments.
%LLMs demand considerably higher computational and storage resources than CNNs due to their large parameter scale and intricate network architectures. Particularly noteworthy is the use of activation functions in Transformers, which are notoriously unfriendly to HE environments.
\item[2)] HE inherently operates through polynomial computations over ciphertexts. Previous encoding strategies inevitably introduce additional computational overhead, such as ciphertext interleaving. Moreover, Bootstrapping, the most time-consuming operation, occurs more frequently as model size increases, further exacerbating computational bottlenecks.
% The complex computational primitives underlying HE substantially increase the computational overhead of ciphertext arithmetic operations, which are invariably much slower than their plaintext counterparts.
\end{itemize}
Consequently, achieving efficient secure inference for LLMs urgently calls for a joint optimization approach encompassing both model architecture design and cryptographic algorithm development.

% In HE-based secure inference for Transformers,the primary bottleneck arises from the extensive matrix multiplication and non-linear activation functions.
%具体来说，在基于HE的LLM安全推理中，主要的瓶颈在于有效地执行广泛的矩阵乘法和评估非线性激活函数。之前的研究{rho2024，Li2024Ni}主要集中在复杂的密文-密文矩阵乘法（CCMM）上，忽略了更简单但更频繁出现的明文-密文矩阵相乘（PCMM）。非线性层的核心挑战在于实现同态不友好操作，特别是softmax函数。现有的方法主要采用高次多项式近似{zhang2024}或HE友好的替代方案{rho2024}。然而，前者具有更高的计算复杂性，而后者需要额外的再训练。此外，之前的编码设计似乎没有充分解决矩阵乘法和后续非线性计算之间的集成问题，导致密文交织操作繁重。这种疏忽引入了额外的转换和通信开销。
Specifically, in HE-based secure inference for LLMs, the primary bottlenecks lie in efficiently performing extensive matrix multiplications and evaluating nonlinear activation functions. Previous research\cite{rho2024, Li2024Ni} has predominantly focused on the complex ciphertext-ciphertext matrix multiplication (CCMM), overlooking the simpler yet more frequent plaintext-ciphertext matrix multiplication (PCMM). The core challenge of non-linear layers lies in implementing homomorphic-unfriendly operations, particularly the softmax function. Existing approaches primarily employ high-degree polynomial approximations\cite{zhang2024} or HE-friendly alternatives\cite{rho2024}. Nevertheless, the former has higher computational complexity, while the latter requires additional retraining. Additionally, previous encoding designs appear to inadequately address the integration between matrix multiplication and subsequent non-linear computations, resulting in heavy ciphertext interleaving operations. This oversight introduces additional conversion and communication overhead.
\subsection{Our Contributions}
In this paper, we propose a non-interactive secure inference framework for LLMs, named ENSI. The framework is built on the principle of co-design, integrating the RNS-CKKS homomorphic encryption scheme\cite{ckks2017, Cheon2019} with a lightweight LLM variant, BitNet\cite{wang2023, Ma2024}. Our main contributions are summarized as follows: 
 
\begin{itemize}
%我们提出了一种协同设计方法，无缝集成加密方案、编码策略和模型优化。具体来说，我们利用列式编码来更好地适应多头自我关注机制。这种方法允许在矩阵乘法后直接执行后续的非线性运算，而不会产生昂贵的密文交织。此外，根据BitNet的结构特性，我们优化了矩阵乘法过程，消除了PCMM中乘法运算的需要，从而显著降低了计算复杂度。
\item \textbf{A Co-Design Secure Inference Framework for LLM.} We propose a co-design framework that seamlessly integrates encryption schemes, encoding strategies, and model optimization. Specifically, we present column-wise encoding to better accommodate the multi-head self-attention mechanism. This approach allows subsequent nonlinear operations to be performed directly after matrix multiplication without incurring costly ciphertext interleaving.
Furthermore, tailored to the structural characteristics of BitNet, we optimize the matrix multiplication process to eliminate the need for multiplication operations within PCMM, thereby significantly reducing computational complexity.

%在HE的背景下，softmax代表了计算成本最高的非线性组件之一。传统方法使用高次多项式近似来近似softmax，导致大量的计算开销。虽然最近的方法用同态友好的替代方案取代了softmax，但它们通常需要额外的模型再训练或微调，从而增加了巨大的资源成本。在在我们的工作中，我们使用sigmoid注意力作为softmax的轻量级且高效的替代方案，该方法无需额外训练即可实现直接替代，同时避免了直接近似通常带来的高计算成本。
\item \textbf{Retraining-free HE implementation for Non-linear function.}  In the context of HE, softmax represents one of the most computationally expensive non-linear components. Conventional methods approximate softmax using high-degree polynomials, resulting in substantial computational overhead. While recent approaches replace softmax with homomorphic-friendly alternatives\cite{rho2024}, they often require additional model retraining or fine-tuning, adding significant resource costs. 
In our work, ENSI employs sigmoid attention \cite{ramapuram2025} as an efficient drop-in replacement for softmax, avoiding the high computational cost of direct approximations.
%In our work, ENSI utilize sigmoid-attention\cite{ramapuram2025} as a lightweight and efficient alternative to softmax, which requires no additional training. This approach enables direct substitution while avoiding the high computational costs typically associated with direct approximation. 

%与最先进的协议NEXUS相比，我们提出的无乘法明文密文矩阵乘法在计算速度上提高了约8倍。在不引入任何额外的预训练的情况下，我们的方法将softmax操作的推理速度提高了2.2到2.6倍。此外，我们的设计最大限度地减少了所需的自举操作次数，实现了现有方案中最低的自举频率。利用我们编码策略的直观性及其跨多个平台的兼容性，我们进一步提供了一种高效的GPU实现。我们的代码库已在[存储库链接]上公开发布。
\item \textbf{Efficient Implementation on CPU and GPU.} Compared to the state-of-the-art protocol NEXUS\cite{zhang2024}, our proposed multiplication-free plaintext-ciphertext matrix multiplication achieves an approximate $5.8\times$ to $8\times$ improvement in computational speed. Without requiring any additional pretraining, our method accelerates inference on softmax by a factor of $2.2\times$ to $2.6\times$. Furthermore, our design minimizes bootstrapping operations, achieving the lowest bootstrapping frequency among existing schemes. Leveraging the intuitiveness of our encoding strategy, we further provide an efficient GPU implementation for matrix multiplication. Our code is open-sourced and accessible at \url{https://github.com/sugarhh/ENSI}
%We intend to fully open-source our implementation\footnote{The link to the repository has been removed for anonymity. We will release the code publicly along with the official version of this work.}. 

\end{itemize}

\section{Related work}
With the widespread adoption of LLMs, secure inference has emerged as a critical research focus. Current investigations primarily follow three main paradigms.
%Current investigations can be broadly classified into two categories: 
%interactive frameworks based on hybrid cryptographic techniques and non-interactive frameworks relying solely on HE.
%随着LLM的广泛采用，安全推理已成为一个关键的研究焦点。目前的研究大致可分为两类：基于混合密码技术的交互式框架和仅依赖HE的非交互式框架。

\textbf{Interactive Secure Inference.} In interactive settings, researchers typically utilize secure multi-party computation independently\cite{li2022mpcformer, dong2023, akimoto2023} or integrate SMPC with homomorphic encryption to achieve an optimal balance between communication overhead and computational efficiency. THE-X\cite{chen2022x} represents a pioneering effort in secure inference with pretrained Transformer models under HE environments. Iron\cite{hao2022} introduces a custom homomorphic encryption protocol to perform secure matrix multiplication, while leveraging SMPC techniques for the secure computation of nonlinear functions in Transformer models. BOLT\cite{pang2024} proposes two encoding strategies, diagonal packing and column-wise packing, to optimize both PCMM and CCMM. BumbleBee\cite{lu2023bu} further customizes effective protocols for nonlinear activation functions, significantly reducing communication costs. 

The core idea behind these approaches is to efficiently execute linear operations using HE while delegating the computation of nonlinear operations to SMPC. However, converting between MPC and HE primitives adds extra complexity, and the significant communication overhead of SMPC protocols further undermines overall system efficiency.
% However, this presents two primary challenges: firstly, the conversion between SMPC and HE primitives introduces additional complexity; secondly, SMPC-based protocols generally entail substantial communication overhead, adversely affecting the overall system efficiency.
%在交互式环境中，研究人员通常独立使用安全多方计算，或将MPC与同态加密集成，以实现通信开销和计算效率之间的最佳平衡。。THE-X{chen2022x}代表了在HE环境下使用预训练变压器模型进行安全推理的开创性努力。Iron{hao2022}引入了一种自定义同态加密协议来执行安全矩阵乘法，同时利用MPC技术来安全计算Transformer模型中的非线性函数。BOLT\cite{pang2024}提出了两种编码策略，对角线打包和列式打包，以优化PCMM和CCMM。大黄蜂{lu2023bu}进一步为非线性激活函数定制了有效的协议，显著降低了通信成本。
%这些方法背后的核心思想是使用HE有效地执行线性运算，同时将非线性运算的计算委托给MPC。然而，这种方法带来了两个主要挑战：首先，MPC和HE原语之间的转换引入了额外的复杂性；其次，基于MPC的协议通常会带来大量的通信开销，对整体系统效率产生不利影响。

\textbf{Non-Interactive Secure Inference.} Given the substantial communication overhead of interactive approaches, HE-based non-interactive secure inference frameworks have gained increasing attention. NEXUS\cite{zhang2024} is the first work to implement a non-interactive secure inference framework for LLMs using HE, enabling the secure evaluation of nonlinear functions such as \textit{Argmax}. Park et al. proposed PowerFormer\cite{park2024}, which replaces the softmax in the attention with their BRP-max function to enable secure inference. Subsequently, Rho et al. proposed an improved HE-friendly LLM architecture that supports personalized fine-tuned inference\cite{rho2024}. Research in this area focuses on minimizing accuracy loss while efficiently performing HE-unfriendly non-linear operations, ensuring that the overall computational cost remains within acceptable limits. Concurrently, for HE protocols, handling a large number of matrix multiplication operations remains a significant challenge, often leading to relatively slow secure inference.
%鉴于交互式方法的大量通信开销，基于HE的非交互式安全推理框架越来越受到关注。NEXUS{zhang2024}是第一个使用HE为变压器实现非交互式安全推理框架的工作，能够对Argmax等非线性函数进行安全评估。Park等人提出了PowerFormer{park2024}，它用BRP max函数替换了注意力中的softmax，以实现安全推理。随后，Rho等人提出了一种改进的HE友好型Transformer架构，支持个性化微调推理{rho2024}。该领域的研究侧重于在有效执行HE不友好的非线性操作的同时，最大限度地减少精度损失，确保整体计算成本保持在可接受的范围内。

\textbf{GPU-based TEEs for Secure Inference.} Unlike schemes that rely on HE or SMPC, GPU-based Trusted Execution Environments (TEEs) establishes a trusted boundary within dedicated hardware, enabling the direct offloading of security-sensitive data to be processed on the GPU\cite{tees-huang2024,tees-li2024teeslice,tees-mohan2024,tees-volos2018g}. While this approach can reduce computational and communication overhead to some extent, it has inherent limitations, such as restricted secure resources and dependence on hardware vendors as the root of trust.

\section{PRELIMINARIES}

\subsection{Notations}
We use $\mathcal{C}$ to represent the client and $\mathcal{S}$ to represent the server. The symbol $\pi$ denotes encoding, $Enc$ and $Dec$ indicate encryption and decryption, respectively. Bold lowercase letters are used to denote vectors, and bold uppercase letters represent matrices. The symbol $\tilde{\textbf{a}}$ refers to a homomorphic ciphertext. Homomorphic addition, subtraction, and multiplication are represented by $\boxplus$, $\boxminus$ and $\boxtimes$, respectively. The rotation of ciphertext is denoted by $Rot$. Regarding the parameters, \(N^\prime\) is the polynomial degree in RNS-CKKS, \(N\) is the number of SIMD slots, which typically satisfies \(N = N^\prime / 2\), and \(L\) indicates the multiplicative depth.

% \subsection{Threat Model}

% Similar to prior work\cite{hao2022, pang2024, lu2023bu, Juvekar2018}, our design is oriented towards the honest-but-curious adversarial model. Although these adversaries follow the protocol specifications, they may attempt to acquire information beyond what the protocol permits. Our protocol remains secure even in the presence of semi-honest adversaries capable of passively compromising either the client or the server.

%与先前的两方计算（2PC）协议类似，我们的设计基于半诚实对手的假设。尽管这些对手遵循协议规范，但他们可能会尝试获取超出协议允许范围的信息。在能够被动破坏客户端或服务器的半诚实对手存在时，我们的协议仍然是安全的。

\subsection{Homomorphic Encryption and RNS-CKKS}
Homomorphic encryption enables computation directly on encrypted data. CKKS \cite{ckks2017} is a homomorphic encryption scheme designed for approximate arithmetic on real and complex numbers, making it well-suited for machine learning tasks. It supports Single Instruction Multiple Data (SIMD) operations by packing multiple values into one single ciphertext. The Residue Number System (RNS) variant, RNS-CKKS \cite{Cheon2019, blatt2020}, is a leveled scheme supporting up to \(L\) multiplicative levels. Both plaintexts and ciphertexts belong to the polynomial ring \(\mathbb{R}_\mathcal{Q} = \mathbb{Z}_\mathcal{Q}[X]/(X^{N^\prime} + 1)\), where \(\mathcal{Q} = \prod_{i=0}^L q_i\) and each \(q_i\) is a distinct prime number.

% Homomorphic encryption allows computation to be performed directly on encrypted data. CKKS \cite{ckks2017} is a homomorphic encryption scheme that enables approximate arithmetic operations on real and complex numbers, making it particularly suitable for handling floating-point numbers and machine learning tasks. By packaging multiple real or complex numbers into a single ciphertext, CKKS supports Single Instruction Multiple Data (SIMD) operations on ciphertexts.
% %同态加密允许直接对加密数据进行计算。CKKS\cite{cheon2017} 是一种同态加密方案，能够对实数和复数进行近似算术运算，使其特别适合处理浮点数和机器学习任务。通过将多个实数或复数打包成一个密文，CKKS支持对密文进行单指令多数据(SIMD)操作。
% The full residue number system (RNS) variant of CKKS\cite{Cheon2019,blatt2020}, RNS-CKKS, is a leveled homomorphic encryption scheme capable of supporting up to $L$ levels of multiplicative depth. Both the plaintext and ciphertexts are elements of a polynomial ring defined as $\mathbb{R}_\mathcal{Q} = \mathbb{Z}_\mathcal{Q}[X]/(X^{N^\prime} + 1)$, where $Q = \prod_{i=0}^{L} q_i$ and $q_i$ are distinct prime numbers.

Bootstrapping (BTS) operations are employed to refresh the level of the ciphertext.
However, due to the substantial computational overhead associated with bootstrapping operation, a meticulous design is essential to minimize this utilization.

%采用自举（BTS）操作来刷新密文的级别。然而，由于与自举操作相关的大量计算开销，细致的设计对于最小化这种利用率至关重要。

The following systematically delineates the fundamental homomorphic operation operators utilized in this paper.

\begin{figure}[!t]\centering
	\includegraphics[width=0.5\textwidth]{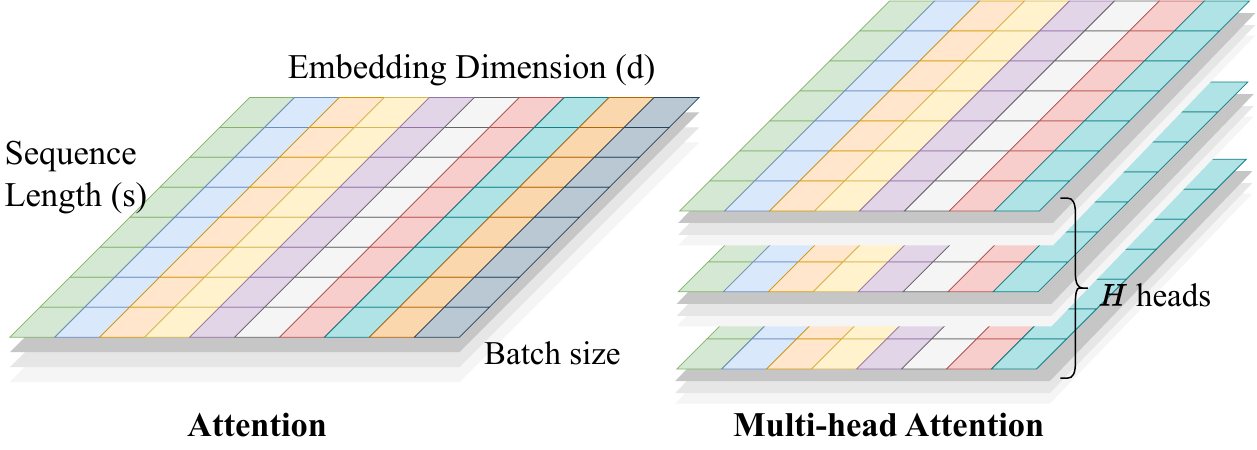}
	\caption{Overview of encoding methods for attention and multi-head attention.}
    \label{encodng}
\end{figure}

\begin{itemize}

\item \textbf{SIMD Encoding}: In previous work \cite{pang2024,rho2024,zhang2024}, matrices are typically encoded into blocks using row or diagonal packing methods. However, these approaches often confine computation within individual ciphertexts or necessitate additional ciphertext operations, which lead to costly ciphertext interleaving. To mitigate these overheads, this paper adopts a direct column-wise encoding strategy, as illustrated in Figure \ref{encodng}. Specifically, $\text{for } j \in [d]$
$$\mathbf{x}_j = \pi(\mathbf{X}_{:, j}) \text{ s.t. } \mathbf{x}_j[i] = \mathbf{X}_{i,j}, \quad\text{ for } i\in [s].$$
Here, $s$ denotes the sequence length and $d$ is the embedding dimension.
Considering that typical LLM sequence lengths are 2048, 4096, or even larger, this packing method efficiently maximizes ciphertext slot utilization while avoiding the need for an excessively large $N^\prime$.

%RNS-CKKS支持单指令多数据（SIMD）操作，允许在单个密文中加密$N$数据向量。在之前的工作中，{pang2024，rho2024，zhang2024}，矩阵通常使用行或对角级联方法编码成块。然而，由于需要在单个密文内进行计算或进行额外的密文调整，因此评估非线性函数变得不那么方便。因此，本文直接采用列式编码策略，如图\ref{encodng}所示。我们采用以下加密方法：。鉴于当前大型模型中的典型序列长度通常为1024、2048甚至更大，这种打包方法有效地最大限度地利用了密文槽，同时避免了使用过大的$N^\prime$。  !!!!!!!确认一下按列编码的英文。

\item \textbf{KeyGen}: Random sampling is employed to generate key pairs, which include the secret key $sk$, public key $pk$, evaluation key $evk$ and others.
%to support homomorphic encryption operations.
%The $KeyGen$ process produces a pair of keys \((sk, pk) \gets KeyGen()\), where \(sk\) is the secret key and \(pk\) is the public key. 

\item \textbf{Encryption}: Given an input vector \(\mathbf{a} \in \mathbb{R}^\ell\), its encryption produces the ciphertext \(\tilde{\mathbf{a}} \gets \mathsf{Enc}_{pk}(\mathbf{a}) \in \mathbb{R}_\mathcal{Q}\).

\item \textbf{Decryption}:  The decryption algorithm transforms a ciphertext \(\tilde{\mathbf{b}} \in \mathbb{R}_\mathcal{Q}\) back into the plaintext vector \(\mathbf{b} \gets \mathsf{Dec}_{sk}(\tilde{\mathbf{b}}) \in \mathbb{R}^\ell\).

\item \textbf{Addition}: Ciphertext-ciphertext addition (\text{Add}) is evaluated as \(\tilde{\mathbf{c}} \gets \tilde{\mathbf{a}} \boxplus \tilde{\mathbf{b}}\). Plaintext-ciphertext addition (\text{pAdd}) is denoted \(\tilde{\mathbf{c}} \gets \tilde{\mathbf{a}} \boxplus \mathbf{b}\).
%Similarly, for subtraction, we have \(\tilde{\mathbf{c}} \gets \tilde{\mathbf{a}} \boxminus \tilde{\mathbf{b}}\) and \(\tilde{\mathbf{c}} \gets \tilde{\mathbf{a}} \boxminus \mathbf{b}\).

\item \textbf{Multiplication}: Ciphertext-ciphertext multiplication (Mult) is computed as \(\tilde{\mathbf{c}} \gets \tilde{\mathbf{a}} \boxtimes \tilde{\mathbf{b}}\). Plaintext-ciphertext multiplication (PMult) is denoted \(\tilde{\mathbf{c}} \gets \tilde{\mathbf{a}} \boxtimes \mathbf{b}\).

\item \textbf{Rotation}:The rotation operation cyclically shifts the ciphertext vector. For an encrypted vector \(\tilde{\mathbf{c}} = [\tilde{c}_0, \tilde{c}_1, \ldots, \tilde{c}_{\ell-1}]\), the rotated vector obtained by \(\tilde{\mathbf{c}}' \gets Rot(\tilde{\mathbf{c}}; k)\) is given by:  
\[
Rot(\tilde{\mathbf{c}}; k) = [\tilde{c}_k, \tilde{c}_{k+1}, \ldots, \tilde{c}_{\ell-1}, \tilde{c}_0, \ldots, \tilde{c}_{k-1}],
\]
where a negative integer \(k\) denotes an inverse rotation.

\end{itemize}

\subsection{Large Language Model LLaMA}
%本文主要研究基于LLaMA的BitNet。LLaMA是一种仅支持解码器的变压器模型，具有多层结构。每一层都由多头自我注意机制和前馈网络（FFN）组成。
In this paper, we focus on BitNet, a large language model architecture built upon the LLaMA.
LLaMA is a decoder-only transformer with a multi-layer structure, where each layer includes a multi-head self-attention mechanism and a feed-forward network (FFN).

\textbf{Input Format.} Each token in the client's input sentence is mapped to a high-dimensional vector. The input is a three-dimensional tensor with dimensions $(batch\_size, seq\_lengths, dims)$. When $batch\_size = 1$, the input $\mathbf{X} \in \mathbb{R}^{s \times d}$ can be viewed as a two-dimensional matrix, where $s$ represents the sequence length (tokens) and $d$ denotes the embedding dimension.
%客户端输入句子中的每个标记都映射到一个高维向量。变压器的输入是三维$（batch \_size，seq\_lengths，dims）$。当$batch \_size$为1时，输入$\mathbf{X}\in\mathbb{R}^{s\times d}$可以被视为一个二维矩阵，其中$s$表示序列长度（标记），$d$表示嵌入维度。

\textbf{Multi-head Attention.} The multi-head attention layer in LLaMA models contextual relationships in the input sequence using linear transformations to produce query, key, and value matrices. Specifically, the input $\mathbf{X} \in \mathbb{R}^{s \times d}$ is projected by parameter matrices $\mathbf{W}^Q, \mathbf{W}^K, \mathbf{W}^V \in \mathbb{R}^{d \times d}$, yielding $\mathbf{Q}, \mathbf{K}, \mathbf{V} \in \mathbb{R}^{s \times d}$. These are then partitioned into $H$ heads, each with dimension $d' = d/H$, and attention scores are computed via scaled dot-product attention:
\[
\text{Attention}(\mathbf{Q}, \mathbf{K}, \mathbf{V}) = \text{Softmax}\left( \frac{\mathbf{Q} \mathbf{K}^\mathrm{T}}{\sqrt{d}} \right) \mathbf{V}.
\]
Outputs from all heads are concatenated and passed through a final linear layer.

\textbf{Feed-Forward Network.} 
FFN is primarily utilized to process the features generated following the self-attention mechanism. Generally, the architecture of an FFN can be represented as follows: 
\begin{equation}
% \text{FFN}(\mathbf{X}) = \text{Activation}(\text{Linear}_1(\mathbf{X}))\cdot\text{Linear}_2(\mathbf{X}),
\text{FFN}(\mathbf{X}) = \text{Linear}_2(\text{Activation}(\text{Linear}_1(\mathbf{X})))
\end{equation}
where $\mathbf{X}$ is the input matrix and $\text{Linear}$ denotes linear transformations. The activation function employed is SwiGLU \cite{shazeer2020}:
\begin{equation}
\text{SwiGLU}(\mathbf{X}) = (\mathbf{X}\mathbf{W}_1) \odot \text{Swish}(\mathbf{X}\mathbf{W}_2),
\end{equation}
where $\odot$ denotes the Hadamard product and the bias $b$ is omitted.
The Swish activation function is commonly implemented as \textbf{SiLU} (Sigmoid Linear Unit), which is mathematically expressed as:
\begin{equation}\label{silu}
\text{SiLU}(\mathbf{X})_{i,j}  = \frac{\mathbf{X}_{i,j} }{1+e^{-\mathbf{X}_{i,j}}}=\mathbf{X}_{i,j}\cdot \sigma(\mathbf{X}_{i,j}),
\end{equation}
where $\sigma$ is the standard sigmoid function.

%前馈神经网络（Feed-Forward Network, FFN）主要用于处理在自注意力机制后生成的特征。通常情况下，FFN 的结构可以表示为：xxx，其中X是输入向量，L1和L2表示线性变换，Activation 是非线性激活函数。LLaMA中使用的激活函数为 SwiGLU，其结构如下：

\textbf{Normalization.} RMSNorm\cite{zhang2024}, an effective normalization technique that stabilizes training by computing the root mean square for each feature channel, is defined as:
% $$\text{RMSNorm}(\mathbf{X}_{ij}) = \frac{\mathbf{X}_{ij}}{\sqrt{\frac{1}{d} \sum_{j=1}^{d} (\mathbf{X}_{ij})^2} + \epsilon} \cdot \gamma,i \in [s], j \in [d]$$
\begin{equation}
\text{RMSNorm}(\mathbf{X})_{i,j} = \frac{\mathbf{X}_{i,j}}{\sqrt{\frac{1}{d} \sum_{k=1}^{d} (\mathbf{X}_{i,k})^2} + \epsilon} \cdot \gamma,  i \in [s],j \in [d]  
\end{equation}
where $\gamma \in \mathbb{R}^d$ is learnable parameter.

\textbf{Positional Encoding.} Rotary Position Embedding (RoPE)\cite{su2024ro} enhances the model’s capacity to encode positional information by applying rotational transformations to the input vectors. The implementation is typically formulated as shown in Equation \ref{rope}. 
\begin{equation} \label{rope}
\mathbf{R}^d_{\Theta,\nu} \mathbf{x}\hspace{-0.2em} = \hspace{-0.4em}
\begin{pmatrix}  
x_0 \\
x_1 \\
x_2 \\
x_3 \\
\vdots \\
x_{d-2} \\
x_{d-1}  
\end{pmatrix} 
\hspace{-0.25em}
\odot  
\hspace{-0.25em}
\begin{pmatrix}  
\cos \nu \theta_0 \\
\cos \nu \theta_0 \\
\cos \nu \theta_1 \\
\cos \nu \theta_1 \\
\vdots \\
\cos \nu \theta_{\frac{d}{2} - 1} \\
\cos \nu \theta_{\frac{d}{2} - 1}  
\end{pmatrix}  
\hspace{-0.25em}
+  \\
\hspace{-0.25em}
\begin{pmatrix}  
-x_1 \\
x_0 \\
-x_3 \\
x_2 \\
\vdots \\
-x_{d-1} \\
x_{d-2}  
\end{pmatrix}  
\hspace{-0.25em}
\odot     
\hspace{-0.25em}
\begin{pmatrix}  
\sin \nu \theta_0 \\
\sin \nu \theta_0 \\
\sin \nu \theta_1 \\
\sin \nu \theta_1 \\
\vdots \\
\sin \nu \theta_{\frac{d}{2} - 1} \\
\sin \nu \theta_{\frac{d}{2} - 1}  
\end{pmatrix}  
\end{equation}
in which the multiplication is conducted element-wise, i.e., each corresponding element is multiplied individually. For a comprehensive explanation of the underlying principles, refer to reference \cite{su2024ro}.

\begin{figure}[!t]\centering
	\includegraphics[width=0.5\textwidth]{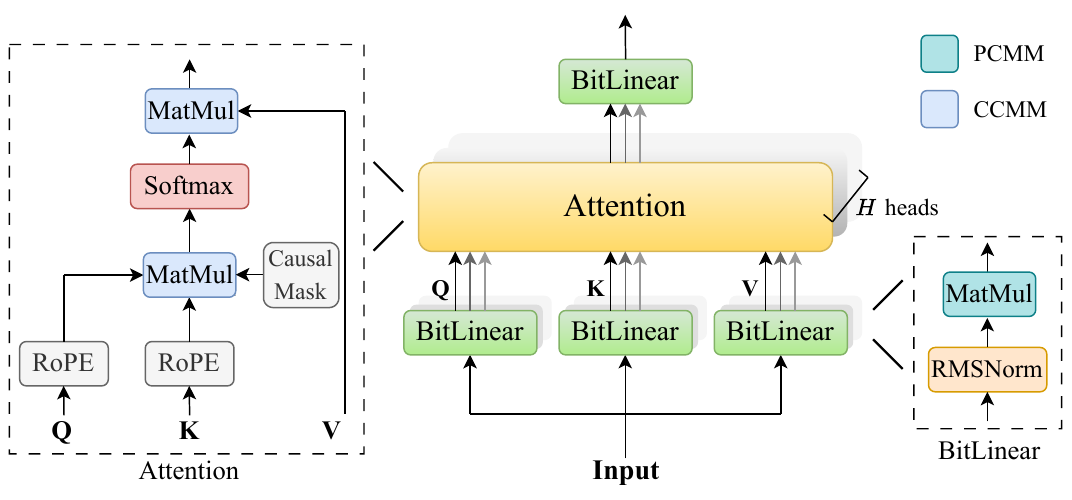}
	\caption{The multi-head attention structure of BitNet}
    \label{att}
\end{figure}

\textbf{BitNet Architecture.} BitNet is a scalable and stable 1-bit large language model architecture designed as an efficient alternative to traditional LLMs such as LLaMA\cite{wang2023}. A notable variant is BitNet b1.58\cite{Ma2024}, where weight parameters are restricted to the ternary set $\left\{ -1, 0, 1\right\}$. 
In the context of the LLaMA architecture, BitNet replaces the standard $nn.Linear$ layers in components such as attention and feed-forward networks with \textit{BitLinear} layers, where outputs are computed as $\mathbf{Y} = \mathbf{W} \cdot \mathbf{X}$ using ternary weight matrices. Notably, since quantization does not affect computation after encryption, this paper does not address quantization related issues. The multi-head attention structure of BitNet is illustrated in Figure~\ref{att}.

% BitNet is a scalable and stable 1-bit LLM architecture designed for LLMs\cite{wang2023}. A notable variant is BitNet b1.58\cite{Ma2024}, where weight parameters are restricted to the ternary set $\left\{ -1, 0, 1\right\}$. Importantly, since quantization does not impact computation speed after encryption, this paper does not consider issues related to quantization. The multi-head attention structure of BitNet is illustrated in Figure \ref{att}.
%BitNet是1位Transformer｛wang2023｝的可扩展且稳定的架构。一个值得注意的变体是BitNet b1.58\cite{Ma2024}，其中权重参数可以从三元集$\left \{-1，0，1\right \}$中取值，并且所有偏差都被消除。重要的是，由于量化在加密后不会影响计算速度，因此本文没有解决与量化相关的问题。BitNet的多头注意力结构如图\ref{att}所示。
% \textbf{BitLinear.} This layer is designed to replace the standard $nn.Linear$ layer. 
% Here, outputs are computed as $\mathbf{Y} = \mathbf{W} \cdot \mathbf{X}$. Unlike standard linear layers, the weight matrix $\mathbf{W} $ is constrained to $ \left\{-1, 0, 1\right\}^{d \times d}$.
% The definition of BitLinear is given by $\mathbf{Y} = \mathbf{W} \cdot \mathbf{X}$. Unlike standard linear layers, the weight matrix $\mathbf{W} \in \left\{-1, 0, 1\right\}^{d \times m}$, where $d$ represents the input dimension and $m$ represents the output dimension. 
%为了实现1位LLM，BitNet采用了一个名为BitLinear的专用层。此层旨在替换标准$nn。线性$layer。BitLinear的定义由$\mathbf{Y}=\mathbf{1W}\cdot\mathbf{0X}$给出。与标准线性层不同，权重矩阵$\mathbf{W}\in\left \{-1，0，1\right \}^{d\times m}$，其中$d$表示输入维度，$m$表示输出维度。值得注意的是，这种方法可以直接消除明文中的乘法运算。然而，如果使用之前的编码方法，仍然需要执行明文-密文矩阵乘法以获得正确的结果。

\section{Problem Statement}
In this work, we aim to enable secure inference for LLMs, allowing users to benefit from powerful server-side models without compromising either the confidentiality of clients’ private data or the proprietary nature of service providers’ model parameters.
%安全推理 允许客户端（用户）在不泄露其私有数据的情况下，利用服务器端的机器学习模型进行推理。图三展示了我们系统的架构。在整个过程中，输入数据的机密性和模型权重与参数的私密性均得到有效保护。
% 客户端： 用户在客户端持有私有数据，并使用同态加密技术根据约定的编码规则对其进行打包和加密。这一过程确保了用户数据在传输和处理中始终处于加密状态，有效防止数据泄露。在推理完成后，服务器将推理结果以加密形式返回给客户端。由于结果也是以加密形式发送，客户端需要进行相应的解密操作以获取可用的推理结果。
% 服务器： 服务器在接收到加密数据后，采用专门设计的安全协议对加密数据进行推理，而无需访问数据的明文内容。出于安全考虑，服务器的模型权重和其他敏感信息不应被客户端获得，因此安全推理机制确保了服务器所持有模型信息的机密性。
\subsection{System Model}
\begin{figure*}[!t]\centering
	\includegraphics[width=0.965\textwidth]{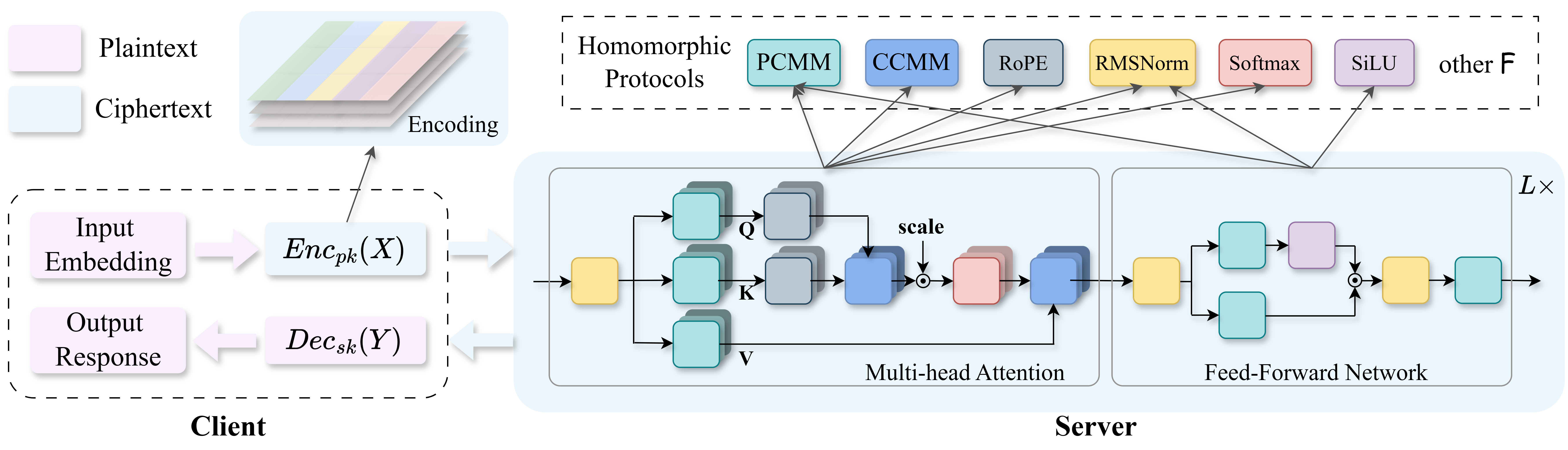}
	\caption{ENSI's high-level architecture.}
    \label{overview}
\end{figure*}

Secure Inference allows clients to utilize machine learning models hosted on servers without disclosing their private data. Figure \ref{overview} illustrates the architecture of our system model. Throughout the process, the confidentiality of the input data, as well as the privacy of the model weights and parameters, is effectively protected.

%图片还要修改

\textbf{Client-Side.} The client holds private data and uses homomorphic encryption techniques to package and encrypt it according to encode algorithm. This process ensures that the user's data remains encrypted during transmission and processing, effectively preventing data leakage. After inference is completed, the server returns the inference results to the client in encrypted form. As the results are also sent in encrypted format, the client must perform the corresponding decryption operations to obtain usable inference results.

\textbf{Server-Side.} Upon receiving the encrypted data, the server employs specially designed secure protocols to conduct inference on the encrypted data without accessing the plaintext content. For security, the model weights and other sensitive information held by the server should not be accessible to the client, thus the secure inference mechanism ensures the confidentiality of the model information possessed by the server.

\subsection{Threat Model}
Similar to prior work\cite{hao2022, pang2024, lu2023bu, Juvekar2018}, our design is oriented towards the honest-but-curious adversarial model. Although these adversaries follow the protocol specifications, they may attempt to acquire information beyond what the protocol permits. Our protocol remains secure even in the presence of semi-honest adversaries capable of passively compromising either the client or the server.

%标题要不要直接写MULTIPLICATION-FREE PCMM
\section{Co-Design with Bitnet: Efficient Secure Linear Computations}
% \section{Bitnet Co-Design: Efficient matrix multiplication}在私有变压器的推理过程中，有两种类型的矩阵乘法（参见图\ref{att}）：（1）密文矩阵和明文矩阵之间的乘法，其中明文矩阵是存储在服务器上的权重矩阵，例如BitLinear中的矩阵乘法；（2）密文矩阵之间的乘法。例如，这发生在注意力机制中查询（$\mathbf{Q}$）和键（$\matbf{K}$）之间的乘法过程中。
% During secure inference with LLMs, there are two main types of matrix multiplication (cf. Figure \ref{att}): (1) multiplication between ciphertext matrices and plaintext matrices, where the plaintext matrix is the weight matrix stored on the server, such as the matrix multiplication in BitLinear; (2) multiplication between ciphertext matrices. For instance, this occurs during the multiplication between query ($\mathbf{Q}$) and key ($\mathbf{K}$) in the attention mechanism.
Linear computations, particularly matrix multiplications, are the primary computational workload in large language models. Under our system setting, there are two main types of matrix multiplications in secure LLM inference (cf. Figure \ref{att}): (1) multiplication between ciphertext matrices and plaintext matrices, where the plaintext matrix is typically the server-side weight matrix (e.g., in \textit{BitLinear}); (2) multiplication between two ciphertext matrices. For instance, this occurs in the attention mechanism during the multiplication between query ($\mathbf{Q}$) and key ($\mathbf{K}$), which both involve client data privacy.

\subsection{Multiplication-Free Plaintext-Ciphertext Matrix Multiplication}

\begin{figure}[ht]\centering
	\includegraphics[width=0.49\textwidth]{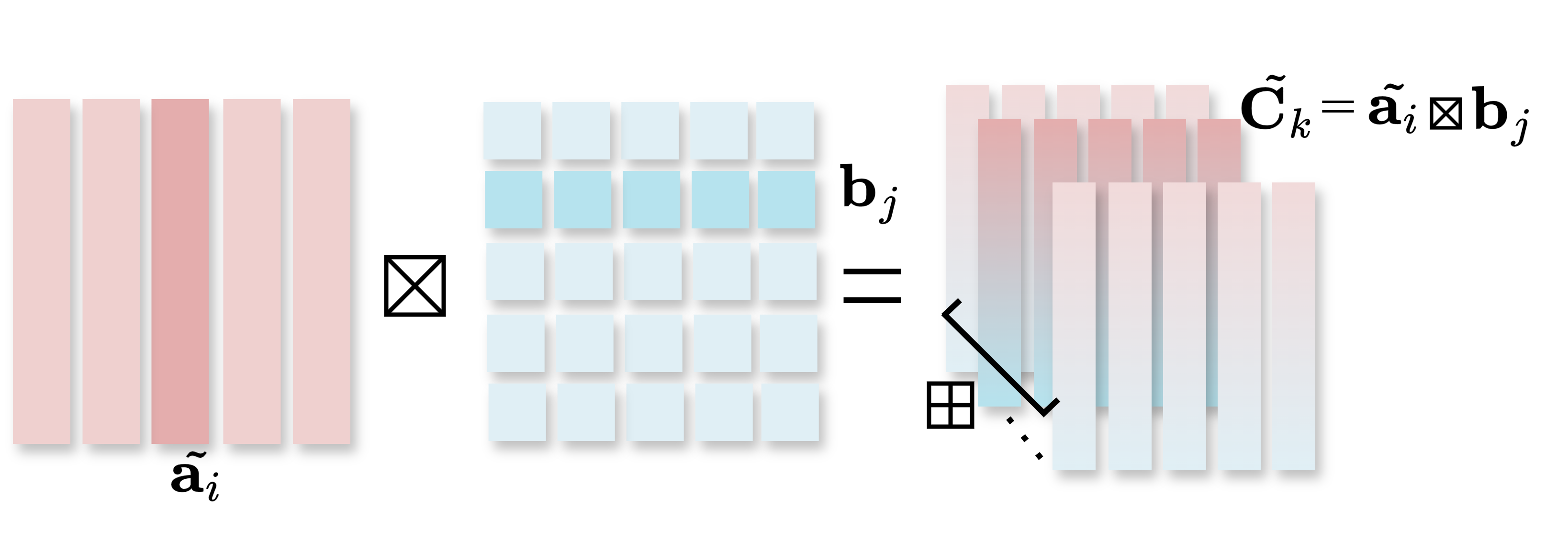}
	\caption{The overview of matrix multiplication based on outer product}
    \label{outmm}
\end{figure}

Matrix multiplication can be viewed as the summation of multiple vector outer products. Specifically, a rank-1 matrix is constructed by the outer product of a column vector and a row vector, and summing these rank-1 matrices results in the final product matrix, as illustrated in Figure \ref{outmm}. Based on this approach, for any ciphertext matrix \(\tilde{\mathbf{A}} \in \mathbb{R}^{s \times d}_\mathcal{Q}\) and plaintext matrix \(\mathbf{B} \in \mathbb{R}^{d \times m}\), the computational complexity involves \(O(dm)\) ciphertext multiplications (pMult) and additions (Add).

Plaintext-ciphertext matrix multiplication is the most commonly used linear computation (see Figure 3), with its efficiency directly impacting overall system performance. To address this bottleneck, we introduce the \textit{BitLinear} structure, which employs a ternary weight matrix $\mathbf{W}\in\{-1, 0, 1\}^{\text{input size} \times \text{output size}}$ to eliminate explicit multiplication in plaintext computation. However, the subsequent challenge is that row or diagonal packing methods fail to deliver computational benefits for ciphertext-level multiplication. In other words, the ciphertext encoding structure limits the exploitation of the algorithm’s advantages, so ciphertext multiplication still requires full-fledged multiplication operations.

% In theory, the use of ternary weights could eliminate explicit multiplication in plaintext matrix computations. However, with standard encoding approaches like row or diagonal packing, its ciphertext-level multiplication offers no advantage over standard matrix multiplication. In other words, the ciphertext encoding structure limits the exploitation of the algorithm’s advantages, so ciphertext multiplication still requires full-fledged multiplication operations.
% Although BitNet’s plaintext matrix multiplication eliminates explicit multiplications, with standard encoding approaches like row or diagonal packing, its ciphertext-level multiplication offers no advantage over standard matrix multiplication. In other words, the ciphertext encoding structure limits the exploitation of the algorithm’s advantages, so ciphertext multiplication still requires full-fledged multiplication operations.

To overcome this limitation, we seamlessly integrate \textit{BitLinear} with our encoding algorithm by employing the co-design approach. We propose the PCMM algorithm, which completely eliminates the need for multiplication operations.
Specifically, given a weight matrix \(\mathbf{W}\) and a ciphertext matrix \(\tilde{\mathbf{X}}\), our objective is to compute the ciphertext output matrix \(\tilde{\mathbf{Y}} = \tilde{\mathbf{X}} \boxtimes \mathbf{W}\), where \(\tilde{\mathbf{X}} \in \mathbb{R}^{s \times d}_\mathcal{Q}\) and \(\mathbf{W} \in \mathbb{R}^{d \times m}\).
% The plaintext matrix corresponding to \(\tilde{\mathbf{X}}\) is \(\mathbf{X} \in \mathbb{R}^{s \times d}\), packaged into multiple column vectors. Accordingly, its ciphertext representation is expressed as \(\tilde{\mathbf{X}} = \left(\tilde{\mathbf{x}}_0, \tilde{\mathbf{x}}_1, \cdots, \tilde{\mathbf{x}}_{d-1}\right)\), where \(\tilde{\mathbf{x}}_i = \{\tilde{x}_{0,i}, \tilde{x}_{1,i}, \cdots, \tilde{x}_{s-1,i}\}\), \(i \in [d]\).
Here, the corresponding plaintext input \(\mathbf{X}\in \mathbb{R}^{s\times d}\) is packed column-wise. Thus, the ciphertext matrix can be expressed as \(\tilde{\mathbf{X}} = (\tilde{\mathbf{x}}_0, \tilde{\mathbf{x}}_1, \cdots, \tilde{\mathbf{x}}_{d-1})\), where \(\tilde{\mathbf{x}}_i = (\tilde{x}_{0,i}, \tilde{x}_{1,i}, \cdots, \tilde{x}_{s-1,i})^\mathrm{T}\) denotes the \(i\)-th column of \(\tilde{\mathbf{X}}\) for \(i \in [d]\).

For a clear illustration of the PCMM algorithm, we present a toy example with parameters \(s = d = 4\) and \(m = 2\). Given the ciphertext matrix \(\tilde{\mathbf{X}} = \left(\tilde{\mathbf{x}}_0, \tilde{\mathbf{x}}_1, \tilde{\mathbf{x}}_2, \tilde{\mathbf{x}}_3\right)\), we compute the resulting ciphertext matrix \(\tilde{\mathbf{Y}}\) as follows:
\[  
\tilde{\mathbf{Y}} = \tilde{\mathbf{X}} \boxtimes \mathbf{W} =  
\left( \tilde{\mathbf{x}}_0, \tilde{\mathbf{x}}_1, \tilde{\mathbf{x}}_2, \tilde{\mathbf{x}}_3 \right) \boxtimes   
\begin{pmatrix}  
1 & -1 \\   
0 & 1 \\   
-1 & 0 \\   
0 & 1   
\end{pmatrix}  
\]
\[
=  
\begin{pmatrix}  
\tilde{\mathbf{x}}_0 \boxminus \tilde{\mathbf{x}}_2,  \tilde{\mathbf{x}}_1 \boxminus \tilde{\mathbf{x}}_0 \boxplus \tilde{\mathbf{x}}_3  
\end{pmatrix}  
=
\begin{pmatrix}  
\tilde{\mathbf{y}}_0 , \tilde{\mathbf{y}}_1  
\end{pmatrix}  
\]  
In summary, for any ciphertext matrix \(\tilde{\mathbf{A}} \in \mathbb{R}^{s \times d}_\mathcal{Q}\) and plaintext matrix \(\mathbf{B} \in \mathbb{R}^{d \times m}\), our PCMM protocol requires only \(O(dm)\) ciphertext additions, completely avoiding the costly homomorphic multiplications. Algorithm \ref{alg:pcmm} provides a detailed description of the proposed PCMM algorithm.
% This algorithm avoids traditional homomorphic multiplication, significantly reducing computational complexity and resource consumption, thus improving the efficiency of ciphertext matrix multiplication.

\begin{algorithm}[h]
    \caption{Secure PCMM protocol, $\Pi_{\textbf{PCMM}}$}
    \label{alg:pcmm}
    \renewcommand{\algorithmicrequire}{\textbf{Input:}}
    \renewcommand{\algorithmicensure}{\textbf{Output:}}
    \begin{algorithmic}[1]
        \REQUIRE  Activation $\tilde{\mathbf{X}} \in \mathbb{R}^{s\times d}_\mathcal{Q} = \left( \tilde{\mathbf{x}}_0, \tilde{\mathbf{x}}_1, \cdots, \tilde{\mathbf{x}}_{d-1}\right)$ and weight $\mathbf{W} \in \mathbb{R}^{d\times m}$ %%input
        \ENSURE PCMM result $\tilde{\mathbf{Y}} \in \mathbb{R}^{s\times m}_\mathcal{Q} = \left( \tilde{\mathbf{y}}_0, \tilde{\mathbf{y}}_1, \cdots, \tilde{\mathbf{y}}_{m-1}\right).$ %%output
        \FOR {$i=0,1,\cdots,m-1$}
            \STATE  $\tilde{\mathbf{y}}_i \gets \left\{ \tilde{y}_{0}, \tilde{y}_{1},\cdots, \tilde{y}_{s-1}\right\}$, where $y_k=0$ and $k \in [s]$
            \FOR {$j=0,1,\cdots,d-1$}
                \IF {$\mathbf{W}_{i,j}=1$}
                    \STATE $\tilde{\mathbf{y}}_i \gets \tilde{\mathbf{y}}_i \boxplus \tilde{\mathbf{x}}_j$
                \ELSIF{$\mathbf{W}_{i,j}=-1$}
                    \STATE $\tilde{\mathbf{y}}_i \gets \tilde{\mathbf{y}}_i \boxminus \tilde{\mathbf{x}}_j$
                \ENDIF
            \ENDFOR 
        \ENDFOR  
        \RETURN $\left( \tilde{\mathbf{y}}_0, \tilde{\mathbf{y}}_1, \cdots, \tilde{\mathbf{y}}_{m-1}\right)$
    \end{algorithmic}
\end{algorithm}

\subsection{Efficient Ciphertext-Ciphertext Matrix Multiplication}
%在下一节中，我们将全面描述基于HE的密文-密文矩阵乘法协议。开发该协议的主要挑战在于通过仔细优化来最小化同态操作的数量，从而减少乘法深度开销。增加的乘法深度加速了密文中的噪声累积，这需要选择更大的HE参数来确保正确的解密。然而，缩放这些参数会带来大量的存储和带宽需求，同时显著降低计算效率。此外，更深的乘法电路会触发更多的自举操作，这本身就会产生巨大的计算成本。
A principal challenge in developing this protocol lies in minimizing the number of homomorphic operations through careful optimization, thereby reducing the multiplicative depth overhead. An increased multiplicative depth accelerates noise accumulation within the ciphertext, which necessitates the selection of larger HE parameters to ensure correct decryption. Moreover, deeper multiplicative circuits trigger frequent bootstrapping procedures, which themselves incur substantial computational costs.

%以前的矩阵乘法方法通常采用行级联编码或对角级联编码方案。这些方法的性能优势主要体现在环参数N足够大时，因为需要更大的N值来有效地支持复杂的编码结构和计算。不幸的是，尽管这些编码策略可以在一定程度上降低计算复杂度，但同时需要大规模HE参数会导致更高的计算开销。
Previous approaches for matrix multiplication commonly employ row-wise packing encoding or diagonal packing encoding schemes. The performance benefits of these methods predominantly manifest when the parameter $N^\prime$ is sufficiently large, as larger values of $N^\prime$ are required to support complex encoding structures and computations effectively. Unfortunately, although these encoding strategies can reduce computational complexity to some extent, the necessity for large-scale HE parameters simultaneously incurs substantially higher computational overhead.

%为了解决注意力机制中固有的矩阵乘法挑战，并利用所提出的编码策略，出现了两种主要形式的密文-密文矩阵乘法：1）列编码矩阵和行编码矩阵之间的乘法（即列编码矩阵的转置）；2） 两个均由列编码的矩阵之间的乘法。给定密文矩阵$\tilde{\mathbf{A}\in\mathbb{R}^{s\times d}_\mathcal{Q}$和$\tilde{\mathbf{B}\in/mathbb{R}^{d\times d'}_\mathcal{Q}$\tilde-{\mathbb{1C}\in\mathbb{R}^{s \times m}_\mathcal{Q}=\tilde{\mathbf{1A}\boxtimes\tilde}\mathbf{B}$，得到的矩阵$\tilte{\mashbf{C}\in \mathbb}^{s\times m}_\mathcal{Q}保留了列编码格式。具体来说，$\tilde{\mathbf{C}}$可以表示为$\tilde{\mathbf{C}=\left（\tilde{\mathbf{1c}_0，\tilde}{\mathbf{C}_1，\ldots，\tilte{\mashbf{C}_{m-1}\right）$，其中每个$\tildi{\mathbf{C}_i$都是通过以下公式计算的：$$\tilde{\mathbf{c}_i}=\tilde{\mathbf{a}_0}\boxtimes \ tilde{\mathbf{B}_｛0，i｝\boxplus\tilde｛\mathbf{a}_1}\boxtimes \ tilde{\mathbf{B}_{1，i}\boxplus \cdots\boxplus \tilde{\mathbf{a}_j}\boxtimes \ tilde{\mathbf{B}_{j，i}}，j\in[d]，i\in[m]$$关键的挑战仍然是如何高效快速地从密文中提取元素$\tilde{\mathbf{B}_{j，i}$，并执行所需的复制操作。
%为了解决注意力机制中固有的矩阵乘法挑战，并利用所提出的编码策略，出现了两种主要形式的密文-密文矩阵乘法：1）列编码矩阵和行编码矩阵之间的乘法（即列编码矩阵的转置）；2） 两个均由列编码的矩阵之间的乘法。基于这两种形式的矩阵乘法的共性，我们开发了一种与这两种类型的矩阵乘法兼容的通用密文-密文矩阵乘法协议。
Addressing the challenges of ciphertext-ciphertext matrix multiplication inherent in the attention mechanism, and by leveraging the proposed encoding strategy, we identify two primary forms: 1) multiplication between a column-encoded matrix and a row-encoded matrix (i.e., the transpose of a column-encoded matrix); 2) multiplication between two matrices both encoded by columns. Based on the commonalities of the two forms of matrix multiplication, we develop a universal ciphertext-ciphertext matrix multiplication protocol that is compatible with both types of matrix multiplication.

\begin{figure}[!t]\centering
	\includegraphics[width=0.5\textwidth]{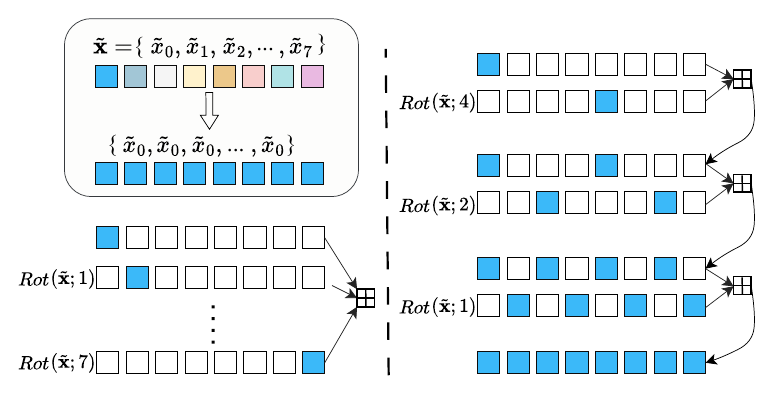}
	\caption{(a) A naive element extraction strategy (left); (b) an innovative extraction method based on BSGS (right).}
    \label{ccmm}
\end{figure}

To illustrate, consider ciphertext matrices $\tilde{\mathbf{A}} \in \mathbb{R}^{s \times d}_\mathcal{Q}$ and $\tilde{\mathbf{B}} \in \mathbb{R}^{d \times m}_\mathcal{Q}$, their product $\tilde{\mathbf{C}} \in \mathbb{R}^{s \times m}_\mathcal{Q} = \tilde{\mathbf{A}} \boxtimes \tilde{\mathbf{B}}$ preserves the column-encoded format. Specifically, $\tilde{\mathbf{C}}$ can be expressed as $\tilde{\mathbf{C}} = \left(\tilde{\mathbf{c}}_0, \tilde{\mathbf{c}}_1, \ldots, \tilde{\mathbf{c}}_{m-1}\right)$, where each $\tilde{\mathbf{c}}_i$ is computed as follows:
$$\tilde{\mathbf{c}}_i = \tilde{\mathbf{a}}_0 \boxtimes \tilde{\mathbf{b}}^{(0)}_i \boxplus \tilde{\mathbf{a}}_1 \boxtimes \tilde{\mathbf{b}}^{(1)}_i \boxplus \cdots \boxplus \tilde{\mathbf{a}}_j \boxtimes \tilde{\mathbf{b}}^{(j)}_i, j\in[d],i\in[m].$$
Here, \(\tilde{\mathbf{b}}^{(j)}_i\) denotes a ciphertext vector composed of repeated copies of \(\mathbf{B}_{j,i}\) packed within a single ciphertext.
Therefore, the core idea of our CCMM algorithm is that each resulting column vector \(\tilde{\mathbf{c}}_i\) is computed by performing ciphertext multiplications (\(\boxtimes\)) between each column vector \(\tilde{\mathbf{a}}_j\) of matrix \(\tilde{\mathbf{A}}\) and the corresponding ciphertext element \(\tilde{\mathbf{B}}_{j,i}\) in matrix \(\tilde{\mathbf{B}}\), followed by ciphertext additions (\(\boxplus\)) to accumulate the results.

% The critical challenge remains how to efficiently and rapidly extract elements $\tilde{\mathbf{B}}_{j,i}$ from ciphertext $\tilde{\mathbf{B}}$ and perform the required replication operations.  
Naturally, the key challenge becomes how to efficiently extract the element \(\tilde{\mathbf{B}}_{j,i}\) from the ciphertext and expand it into a ciphertext vector of length \(s\). In CKKS, directly extracting a single element from a ciphertext is nearly impossible. The most straightforward extraction strategy relies on multiplying by a mask vector, (e.g.,\(\{1,0,0,\cdots,0\}\))combined with multiple rotations to achieve element extraction and position alignment, as illustrated in Figure \ref{ccmm}(a). This approach requires approximately $O(dm)$ ciphertext multiplications, rotations, and additions, resulting in a significant computational burden.

To overcome this inefficiency, we draw inspiration from the baby-step giant-step (BSGS) algorithm\cite{basu2014baby} and introduce an innovative extraction mechanism, shown in Figure \(\ref{ccmm}\) (b). This novel method effectively reduces the number of rotations from linear to logarithmic scale, specifically down to \ $\log(dm)$. This optimization significantly reduces the overall rotation overhead. 
It is worth noting that the independence and similarity of operations across different ciphertexts make this scheme especially well-suited for parallel computing, significantly improving overall computational efficiency.

%我们提出了一种简单而高效的元素提取算法，适用于行编码和列编码矩阵。如图\ref{ccmm}（a）所示，直接的提取策略依赖于乘以掩码向量$\{1,0,0，\cdots，0\}$，并结合多次旋转来实现元素提取和位置对齐。这种方法需要大约$dm$pMult、Rot和Add，导致了巨大的计算负担。为了解决这种低效问题，我们从婴儿步大步（BGSG）算法中汲取灵感，并引入了一种创新的提取机制，如图\ref{ccmm}（b）所示。这种新颖的方法有效地将旋转次数从线性减少到对数尺度，特别是减少到$\log（dm）$。这种优化显著降低了整体旋转开销。
% We propose a simple yet efficient element extraction algorithm applicable to both row-encoded and column-encoded matrices. As illustrated in Figure \ref{ccmm} (a), a straightforward extraction strategy relies on multiplying by a mask vector $\{1,0,0,\cdots,0\}$, combined with multiple rotations to achieve element extraction and positional alignment. This approach requires approximately $O(dm)$ pMult, Rot and Add, resulting in a major computational burden. To address this inefficiency, we draw inspiration from the Baby-step giant-step (BGSG) algorithm and introduce an innovative extraction mechanism depicted in Figure \ref{ccmm} (b). This novel approach effectively reduces the number of rotations from linear to logarithmic scale, specifically to $\log(dm)$. This optimization significantly diminishes the overall rotation overhead. 
% %尽管优化的提取策略仍然需要增加计算复杂性，但不同密文之间操作的独立性使批处理技术的应用能够并行执行计算。这种并行性大大提高了整体计算效率。
% Although the optimized extraction strategy still entails an increased computational complexity, the independence of operations across different ciphertexts enables the application of batching techniques to execute computations in parallel. This parallelism substantially enhances the overall computational efficiency.  

\subsection{Positional Encoding On Ciphertext}
% Besides matrix multiplication, linear operations on ciphertexts also include RoPE and Causal-mask functions.
%RoPE将位置信息编码为向量空间中的旋转变换，从而能够连续无偏地表示序列中元素的相对位置。由于RoPE中使用的$cos$和$sin$向量可以预先计算，因此本文不讨论它们的生成过程；相反，我们直接应用它们。方程式\ref{rope}说明了主要的计算过程。RoPE以相同的方式应用于查询（$\mathbf{Q}$）和关键字（$\matbf{K}$）向量。算法\ref{alg:RoPE}详细说明了如何对加密的$\mathbf{Q}$安全地执行位置编码。

% \textbf{Secure RoPE Evaluation.} 
Apart from matrix multiplication, linear operations on ciphertexts also encompass RoPE. RoPE encodes positional information as rotational transformations in the vector space, enabling continuous and unbiased representation of the relative positions of elements in a sequence. The $cos$ and $sin$ vectors used in RoPE can be precomputed, so their generation process is not discussed in this paper and they are applied directly. Equation \ref{rope} illustrates the main computational procedure. RoPE is applied to both the query ($\mathbf{Q}$) and key ($\mathbf{K}$) vectors in an identical manner. The algorithm \ref{alg:RoPE} details how to securely perform positional encoding on the encrypted $\mathbf{Q}$.

\begin{algorithm}[!h]
    \caption{Secure RoPE protocol, $\Pi_{\textbf{RoPE}}$}
    \label{alg:RoPE}
    \renewcommand{\algorithmicrequire}{\textbf{Input:}}
    \renewcommand{\algorithmicensure}{\textbf{Output:}}
    \begin{algorithmic}[1]
        \REQUIRE $\tilde{\mathbf{Q}} \in \mathbb{R}^{s\times d}_\mathcal{Q} = \left( \tilde{\mathbf{x}}_0, \tilde{\mathbf{x}}_1, \cdots, \tilde{\mathbf{x}}_{d-1}\right)$ and $\text{sin}$, $\text{cos}$ vectors%%input
        \ENSURE RoPE result $\tilde{\mathbf{Y}} \in \mathbb{R}^{s\times d}_\mathcal{Q} = \left( \tilde{\mathbf{y}}_0, \tilde{\mathbf{y}}_1, \cdots, \tilde{\mathbf{y}}_{d-1}\right).$ %%output
        \STATE Let $\mathbf{neg} = \left\{-1, 0,-1,0, \cdots,-1,0\right\}$ and $\mathbf{pos} = \left\{ 0,1,0,1 \cdots, 0, 1\right\}$
        \FOR {$i=0,1,\cdots,d-1$}
            % \STATE $\tilde{\mathbf{r}}_i \gets Rot(\tilde{\mathbf{q}_i}; 1)$ 
            \STATE $\tilde{\mathbf{r}}_i \gets Rot(\tilde{\mathbf{q}_i}; 1)\boxtimes \mathbf{neg}$ 
            % \STATE $\tilde{\mathbf{l}}_i \gets Rot(\tilde{\mathbf{q}_i}; -1)$ 
            \STATE $\tilde{\mathbf{l}}_i \gets Rot(\tilde{\mathbf{q}_i}; -1) \boxtimes \mathbf{pos}$ 
            \STATE $\tilde{\mathbf{t}}_i \gets \tilde{\mathbf{r}}_i \boxplus \tilde{\mathbf{l}}_i$  \COMMENT{\textsc{rotational sign exchange}}
            \STATE $\tilde{\mathbf{y}}_i \gets (\tilde{\mathbf{q}}_i \boxtimes \text{cos})\boxplus (\tilde{\mathbf{t}}_i \boxtimes \text{sin})$  
        \ENDFOR 
        \RETURN $\left( \tilde{\mathbf{y}}_0, \tilde{\mathbf{y}}_1, \cdots, \tilde{\mathbf{y}}_{d-1}\right)$
    \end{algorithmic}
\end{algorithm}

% \textbf{Secure Causal-mask Evaluation.} The causal mask in the linear module can be efficiently implemented using additive operations. Specifically, these operations are performed using the basic operator pAdd. The secure causal-mask process is not elaborated further in this paper.

\section{Co-Design with CKKS: Optimizing Non-Linear Operations}
% \section{Co-Design with CKKS:  Efficient Non-Linear Layers}
% 由于密文上非线性函数的同态估计通常需要大量的运算，因此会导致快速的噪声累积和较高的计算开销。这使得非线性函数通常被认为是同态不友好的，并且是构建具有同态加密的安全推理系统的主要瓶颈。
As the homomorphic evaluation of nonlinear functions on ciphertexts typically requires numerous operations, it leads to rapid noise accumulation and high computational overhead. This makes nonlinear functions generally considered HE-unfriendly and represents a major bottleneck in building secure inference systems with homomorphic encryption.
\subsection{Secure Softmax Evaluation Without Retraining}
%softmax函数构成了整个推理过程中的主要性能瓶颈。先前的研究主要采用安全多方计算技术，通过分段多项式函数近似非线性函数。虽然有效，但这些方法会因执行比较操作而产生大量的通信开销。Rho等人提出利用高斯核（GK）方法完全绕过直接softmax计算，从而提高推理效率。然而，这种方法的有效性在很大程度上取决于密文约束下的模型再训练，这需要大量的计算成本，并限制了实际应用。因此，确定一个可以直接替换softmax而不会产生昂贵的再训练开销的替代函数仍然是一个理想的目标。
The softmax function constitutes a primary performance bottleneck throughout the inference process. Prior studies\cite{lu2023bu, pang2024,Li2024Ni} predominantly adopt secure multiparty computation techniques, approximating nonlinear functions via piecewise polynomial functions. Although effective, these approaches incur substantial communication overhead arising from the implementation of comparison operations. Rho et al. proposed leveraging Gaussian Kernel (GK) methods to entirely circumvent direct softmax computation\cite{rho2024}, thereby improving inference efficiency. However, this method’s efficacy depends heavily on model retraining under ciphertext constraints, which entails significant computational cost and restricts practical applicability. Consequently, identifying a substitute function that can directly replace softmax without incurring expensive retraining overhead remains an open challenge.

%利用大型语言模型的快速发展，Ramapuram等人引入了Sigmoid Attention机制{ramapuram2025}
Capitalizing on the rapid advances in large language models, Ramapuram et al. introduced the Sigmoid Attention mechanism\cite{ramapuram2025}:
\begin{equation}
    \text{SigmoidAttn}(\mathbf{X}) = \sigma\left(\frac{\mathbf{Q} \mathbf{K}^\mathrm{T}}{\sqrt{d}}\right) \mathbf{V},  
\end{equation}
$$
\text{with } \sigma : u \mapsto \text{sigmoid}(u + b) := \left(1 + e^{-(u+b)}\right)^{-1}.  
$$
In the above equation, the bias term is defined as $b= \log(s)$
Similar to previous work\cite{rho2024,zhang2024}, the need to calculate exponential functions has not been eliminated. But compared to softmax, sigmoid possesses considerably lower computational complexity. This characteristic renders sigmoid more feasible and efficient for homomorphic encryption environments.
It is also worth noting that softmax is typically applied to each row to normalize the weights into a probability distribution. However, with common encoding strategies such as row packing or diagonal packing, this either introduces additional ciphertext permutations or requires intra-ciphertext operations. In contrast, our encoding strategy avoids both ciphertext permutations and intra-ciphertext operations, offering significant advantages.
%在上述方程式中，偏差项定义为$b=\log（s）$与之前的工作类似，计算指数函数的需要并没有被消除。但与softmax相比，sigmoid的计算复杂度要低得多。这一特性使得sigmoid在同态加密环境中更可行、更高效。

For ciphertext evaluation of the sigmoid function, we employ Chebyshev polynomial approximation combined with the Paterson-Stockmeyer algorithm \cite{al2022openfhe} to achieve numerically stable and efficient polynomial evaluation. Chebyshev polynomials is the preferred method for approximating smooth functions, owing to their minimax property that minimizes the maximum approximation error on the interval $[-1, 1]$.

Specifically, given an input interval $[a,b]$ and a polynomial approximation degree $n$, increasing the degree improves the approximation accuracy but also elevates the computational complexity. The process begins with computing $K = n+1$ Chebyshev nodes within the canonical interval $[-1, 1]$:
$$x_j = \cos\left(\frac{\pi (j + 0.5)}{K}\right), \quad j = 0,1,\ldots, K-1.$$
Convert nodes to the target $[a,b]$ interval through linear mapping $t_j = \frac{b - a}{2} x_j + \frac{a + b}{2}.$ Function values are then evaluated at these mapped Chebyshev nodes, denoted as $f_j=\varsigma (t_j),$ and the coefficients are computed using the following formula: $$c_i = \frac{2}{K} \sum_{j=0}^{K-1} f_j \cos\left(i \cdot \frac{\pi (j + 0.5)}{K}\right), \quad i = 0,1,\ldots, K-1.$$ This procedure yields the corresponding coefficient set $\left\{c_0, c_1, \cdots, c_{K-1} \right\}$. 
Finally, the corresponding polynomial value is evaluated according to the definition of the Chebyshev polynomial:
\begin{equation}
P(\tilde{t}) = \sum_{i=0}^{n} c_i T_i \left(  \frac{2 \cdot \tilde{t} - (b + a)}{b - a} \right).
\end{equation}

%Paterson-Stockmeyer算法通过将多项式分解为块来减少所需的乘法次数。直接计算一个阶-$n$多项式需要大约$n$次乘法，而Paterson-Stockmeyer方法将其减少到大约$\sqrt{n}$次乘法。由于同态加密中的乘法运算在计算上很昂贵，并且电路深度有限，因此减少乘法次数可以显著提高效率和可行性。
The Paterson-Stockmeyer algorithm reduces the number of multiplications required by decomposing the polynomial into blocks. Direct evaluation of a degree-$n$ polynomial requires approximately $n$ multiplications, whereas the Paterson-Stockmeyer method reduces this to roughly $\sqrt{n}$ multiplications. Since multiplication operations in homomorphic encryption are computationally expensive and have limited circuit depth, reducing the number of multiplications can significantly improve both efficiency and feasibility.

\subsection{Normalization and Activation Functions on ciphertext}
\textbf{Secure RMSNorm Evaluation.} RMSNorm is a computationally complex normalization method, and its secure implementation becomes challenging due to the difficulty of performing division and square root operations under HE. We utilize Chebyshev polynomials to efficiently approximate these operations ($\Pi_{SQRT}$, $\Pi_{INVERSE}$), enabling secure computation. Algorithm \ref{alg:rms} provides a detailed description of the secure implementation procedure for RMSNorm.
%RMSNorm是一种计算复杂的归一化方法，由于在HE下执行除法和平方根运算的困难，其安全实现面临着重大挑战。我们建议使用切比雪夫多项式近似来有效地近似这些操作，从而实现安全计算。Algorithm\ref{rms}提供了RMSNorm安全实现过程的详细描述。
\begin{algorithm}[!h]
    \caption{Secure RMSNorm protocol, $\Pi_{\textbf{RMSNorm}}$}
    \label{alg:rms}
    \renewcommand{\algorithmicrequire}{\textbf{Input:}}
    \renewcommand{\algorithmicensure}{\textbf{Output:}}
    \begin{algorithmic}[1]
        \REQUIRE  Activation $\tilde{\mathbf{X}} \in \mathbb{R}^{s\times d}_\mathcal{Q} = \left( \tilde{\mathbf{x}}_0, \tilde{\mathbf{x}}_1, \cdots, \tilde{\mathbf{x}}_{d-1}\right)$ and parameter $\gamma \in \mathbb{R}^d$ %%input
        \ENSURE RMSNorm result $\tilde{\mathbf{Y}} \in \mathbb{R}^{s\times d}_\mathcal{Q} = \left( \tilde{\mathbf{y}}_0, \tilde{\mathbf{y}}_1, \cdots, \tilde{\mathbf{y}}_{d-1}\right).$ %%output
        \FOR {$i=0,1,\cdots,d-1$}
            \STATE $\tilde{\mathbf{t}}_i \gets \tilde{\mathbf{x}}_i \boxtimes \tilde{\mathbf{x}}_i$ \COMMENT{\textsc{square}}
            \STATE $\tilde{\mathbf{s}} \gets \tilde{\mathbf{s}} \boxplus \tilde{\mathbf{t}}_i$ \COMMENT{\textsc{sum}}
        \ENDFOR 
        \STATE $\mathbf{inv} = \left\{\frac{1}{d}, \frac{1}{d},\cdots,\frac{1}{d}\right\}$
        \STATE $\tilde{\mathbf{s}} \gets \tilde{\mathbf{s}} \boxtimes \mathbf{inv}$ \COMMENT{\textsc{variance}}
        \STATE $\tilde{\mathbf{s}} \gets \Pi_{\text{SQRT}}(\tilde{\mathbf{s}})$ \COMMENT{\textsc{square root}}
        \STATE $\tilde{\mathbf{s}} \gets \Pi_{\text{INVERSE}}(\tilde{\mathbf{s}})$ \COMMENT{\textsc{inverse}}
        \FOR {$i=0,1,\cdots,d-1$}
            \STATE $\tilde{\mathbf{y}}_i \gets \tilde{\mathbf{x}}_i \boxtimes \tilde{\mathbf{s}}$ 
            \STATE $\tilde{\mathbf{y}}_i \gets \tilde{\mathbf{y_i}} \boxtimes\gamma$ 
        \ENDFOR   
        \RETURN $\left( \tilde{\mathbf{y}}_0, \tilde{\mathbf{y}}_1, \cdots, \tilde{\mathbf{y}}_{d-1}\right)$
    \end{algorithmic}
\end{algorithm}

It is important to note that RMSNorm performs normalization along the last dimension of the input.
Previous encryption schemes, although based on polynomial approximations, were constrained by their encoding strategies, such as row or diagonal packing, which necessitated executing complex operations within the encrypted ciphertext vector. It is well established that performing homomorphic computations internally on the ciphertext introduces considerable computational complexity. In contrast, our proposed column-vector packing encoding strategy is both more straightforward and efficient. The principal advantage of our approach lies in eliminating internal vector operations such as summation and alignment present in prior methods, thereby confining all computations to the vector level. This design significantly reduces computational overhead and completely removes the need for additional ciphertext manipulation steps.
%以前的加密方案虽然基于多项式近似，但受到其编码策略的限制，如行或对角线打包，这需要在加密的密文向量内执行复杂的操作。众所周知，在密文内部执行同态计算会引入相当大的计算复杂性。相比之下，我们提出的列向量打包编码策略既简单又高效。我们方法的主要优点在于消除了现有方法中存在的求和和和对齐等内部向量操作，从而将所有计算限制在向量级别。这种设计显著降低了计算开销，完全消除了对额外密文操作步骤的需要。

Building on our proposed encoding design, we observed that during the RMSNorm computation, the ciphertext dimension is compressed to 1 (Steps 5 to 8 in Algorithm \ref{alg:rms}). This property provides a key insight for designing a minimized bootstrapping operation, thereby significantly reducing the computational overhead of bootstrapping.

% Moreover, we observe that prior work introduces a Dynamic Tanh function (DyT)\cite{zhu2025trans}, defined as: 
% \begin{equation}
%    \text{DyT}(\mathbf{X})_{i,j} = \gamma_j \cdot \tanh(\alpha \mathbf{X}_{i,j}) + \beta_j, \quad i \in [s], j \in [d]   
% \end{equation}
% Here, $\alpha$ is a learnable scalar parameter that dynamically adjusts the scaling according to the input range. $\gamma_j$ and $\beta_j$ are learnable vector parameters for each channel, consistent with parameters used in conventional normalization layers, allowing output scaling over arbitrary ranges. DyT can simulate the behavior of Layer Normalization without computing the activation statistics. It applies a nonlinear transformation to the input via the tanh function while preserving the compression effect of normalization layers on extreme values. Although Dynamic requires retraining, empirical results demonstrate that it effectively improves inference speed. Therefore, we consider it a viable option when retraining is feasible.

% Notably, the tanh function can be expressed as a composition of sigmoid functions, specifically 
% $$\tanh(x) = 2 \cdot \sigma(2x) - 1.$$
% Therefore, we can directly leverage the previously implemented sigmoid function to securely evaluate DyT.

\textbf{Secure SiLU Evaluation.} Finally, we discuss the SiLU activation function. In secure computation environments, the implementation of SiLU primarily depends on the computation of the sigmoid function, as shown in Equation \ref{silu}. We can securely evaluate SiLU by leveraging the sigmoid secure inference algorithm used in softmax, combined with multiplication operations. This approach simplifies the secure computation of the activation function and improves overall implementation efficiency.
%最后，我们讨论了SiLU激活函数。在安全计算环境中，SiLU的实现主要取决于sigmoid函数的计算，如方程1所示。我们可以通过利用softmax中使用的sigmoid安全推理算法并结合乘法运算来安全地评估SiLU。这种方法简化了激活函数的安全计算，提高了整体实现效率。

\section{BTS for Minimal Implementation}
% \section{Adoption of bootstrapping}
%由于RNS-CKKS是一种分级同态加密方案，支持每个密文$L$层的最大乘法深度，因此刷新密文级别的自举操作变得不可避免。然而，自举带来了巨大的计算开销，其成本随着输入密文的数量呈线性增长，在实际应用中带来了显著的性能挑战。重要的是，执行BTS会消耗一定数量的级别，表示为$K$，将有效可用级别从最初的$L$降低到$L-K$。一旦所有级别都耗尽，进一步的同态操作就变得不可行，因为密文中的噪声变得太大，无法进行正确的解密。BTS通过评估解密电路和应用模切换技术来缓解这个问题，有效地降低了密文噪声，从而扩展了可行同态计算的深度。
The bootstrapping operation empowers the RNS-CKKS encryption scheme to support homomorphic computations of arbitrary depth. However, bootstrapping introduces significant computational overhead that increases linearly with the number of input ciphertexts, creating notable performance challenges in practice. 
% The bootstrapping operation enables the RNS-CKKS scheme to support homomorphic computations of arbitrary depth, but it incurs significant computational overhead that grows linearly with the number of input ciphertexts, affecting system performance.
Importantly, performing bootstrapping consumes a certain number of levels, denoted as $K$, reducing the effective available levels from the initial $L$ to $L - K$. 
%Once all levels are depleted, further homomorphic operations become infeasible as the noise within the ciphertext becomes too large to allow correct decryption. 

%BTS mitigates this issue by evaluating decryption circuits and applying modulus switching techniques, thereby extending the homomorphic computation depth with full homomorphic properties.

\begin{figure}[!t]\centering
	\includegraphics[width=0.42\textwidth]{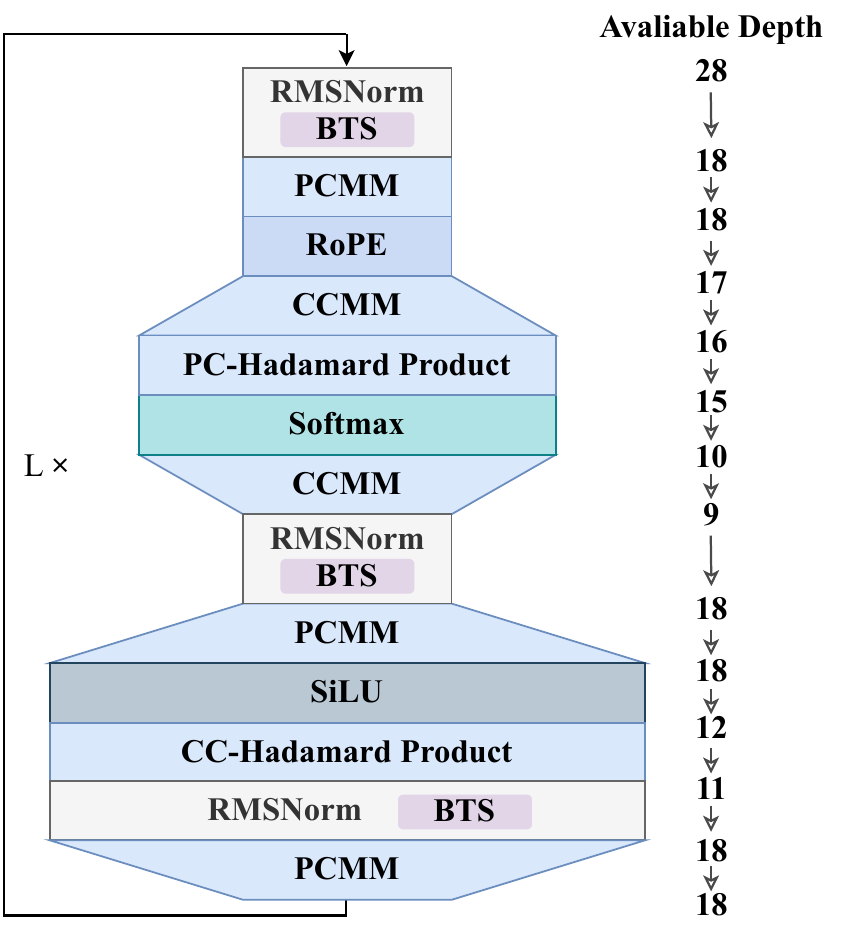}
	\caption{Implementation of bootstrapping in a Bitnet-based architecture, exemplified by LLaMA.}
    \label{fbts}
\end{figure}

Figure \ref{fbts} provides a detailed illustration of the level consumption in our design architecture and the specific position of the bootstrapping operation. 
% In our design framework, the RMSNorm layer, as one of the most computation-intensive layers, consumes up to 19 multiplication levels. Specifically, the initial 7 steps of Algorithm \ref{alg:rms} cumulatively account for 9 multiplicative levels, while the subsequent operations consume 10 levels.
The RMSNorm layer, the most computation-intensive in our design, consumes 19 multiplication levels (9 levels in the first 7 algorithm steps and 10 in the subsequent ones).
Based on this, we observed that performing bootstrapping at any point between Steps 6 and 8 of Algorithm \ref{alg:rms} can substantially reduce the frequency of bootstrapping operations. Precisely, by executing the bootstrapping procedure immediately after RMSNorm has consumed 9 multiplicative levels (i.e., after the $\Pi_{\text{SQRT}}$), we are able to effectively refresh the ciphertext noise. For any RMSNorm input matrices $\tilde{\mathbf{X}} \in \mathbb{R}^{s\times d}_\mathcal{Q}$, our design reduces the complexity of bootstrapping from the original $O(d)$ to $O(1)$. This approach results in the lowest bootstrapping overhead observed within the current framework.

% Based on this, we observed that performing bootstrapping immediately after the $\Pi_{\text{SQRT}}$ operation in RMSNorm significantly reduces the frequency of bootstrapping. Specifically, after RMSNorm consumes 18 levels, a bootstrapping operation is executed to refresh ciphertext noise before continuing subsequent multiplications. For any RMSNorm input matrices $\tilde{\mathbf{X}} \in \mathbb{R}^{s\times d}_\mathcal{Q}$, our design reduces the complexity of bootstrapping from the original $O(d)$ to $O(1)$.

%image

\section{Evaluation}

\subsection{Implementation}
%In this study, we implemented ENSI in Python using the OpenFHE library \cite{al2022openfhe} to enable homomorphic encryption and bootstrapping operations with the CKKS scheme. Owing to the straightforward linear layer architecture—designed without complex packing or transformation steps—our approach maintains strong compatibility with a variety of homomorphic encryption frameworks. To further enhance computational efficiency, we integrated GPU-accelerated operations through the Phantom library \cite{Phantom}, which is developed in C++. For the performance-critical linear layer, we also incorporated the Microsoft SEAL library \cite{sealcrypto}, thereby demonstrating the portability of our implementation and facilitating a fair comparative evaluation.
In this study, we implemented the ENSI in Python using the OpenFHE library\cite{al2022openfhe} to enable homomorphic encryption and bootstrapping operations with the CKKS scheme. Owing to the linear layer architecture designed without complex packing and transformation procedures, our approach exhibits strong compatibility with various homomorphic encryption frameworks. To further enhance computational efficiency, we integrated GPU-accelerated operations through the Phantom library \cite{Phantom}, which is developed in C++. For the performance-critical linear layer, we further incorporated the Microsoft SEAL library\cite{sealcrypto} to demonstrate the transferability of our implementation and to enable a fair comparative evaluation. 
%Moreover, we integrated GPU-accelerated computations via the Phantom library\cite{Phantom}, developed in C++, to further optimize efficiency.

Regarding the cryptographic parameters, all selections were made to comply with a 128-bit security level according to established Homomorphic Encryption Standard\cite{Standard}. During experimentation, parameters with $N^\prime=14$ and $N^\prime=16$ were applied. It is notable that smaller values of $N^\prime$ correspond to faster homomorphic operations; unless otherwise specified, $N^\prime=14$ was employed as the default setting. The multiplicative circuit depth was configured as $L = 48$, while the bootstrapping operation requires a depth of $K=20$. Consequently, the effective supported multiplicative depth of the system is $L - K = 28$.

\subsection{Experimental setup}
%我们对基于同态加密（HE）的两个最具代表性的非交互式安全推理框架进行了比较分析：NEXUS\cite{zhang2024}，它目前代表了该领域的最新技术，以及\cite{rho2024}中提出的框架，为清楚起见，以下称为EFLA。利用他们的开源实现，我们在相同的实验设置下进行了基准评估。对于CPU基准测试，实验是在32核的Intel Core i9-14900K处理器上进行的。然而，鉴于OpenFHE中的多线程并行性仍在开发中，所有实验都是在单核模式下执行的，以确保一致性。对于GPU基准测试，使用了具有80GB内存的Tesla A100 GPU。所有报告的结果代表了10次独立运行的平均值。
We performed a comparative analysis against the two leading non-interactive secure inference frameworks based on HE: NEXUS\cite{zhang2024}, which currently represents the state-of-the-art in this field, and the framework proposed in \cite{rho2024}, hereafter denoted as EFLA for clarity.
Utilizing their open-source implementations, we performed benchmark evaluations under identical experimental settings. For the CPU benchmarks, experiments were conducted on an Intel Core i9-14900K processor with 32 cores. For the GPU benchmarks, a Tesla A100 GPU with 80GB of memory was employed. All reported results represent the averages over 10 independent runs.

%需要强调的是，我们的协议是非交互式的，通信仅限于数据传输和接收。此外，我们的编码设计实现了近乎完全的时隙利用率，导致通信开销和数据量之间存在线性相关性。因此，模拟局域网或广域网环境下的网络通信意义有限。本研究专门关注评估服务器端安全推理协议的运行时效率。
It is necessary to emphasize that our protocol is non-interactive, with communication limited solely to data transmission and reception. Moreover, our encoding design achieves near-complete slot utilization, resulting in a linear correlation between communication overhead and data volume. Therefore, simulating network communication over LAN or WAN environments holds limited significance. This study focuses exclusively on evaluating the runtime efficiency of the secure inference protocol at the server side.

\subsection{Microbenchmarks}
\begin{figure}[!t]\centering
	\includegraphics[width=0.5\textwidth]{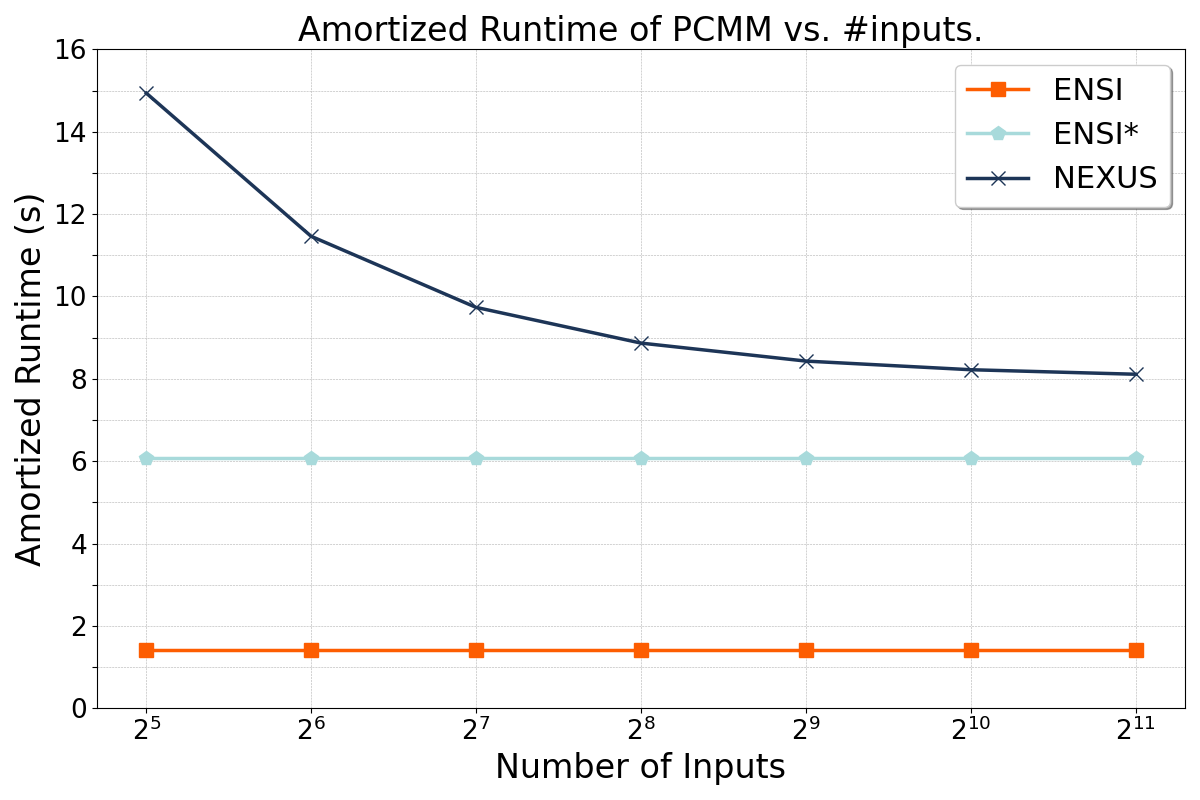}
	\caption{Evaluation of runtime for $\mathbb{R}^{2048\times768}\boxtimes \mathbb{R}^{768\times64}$ ciphertext-plaintext matrix multiplication.}
    \label{time_pcmm}
\end{figure}

%在NEXUS的明文-密文矩阵乘法协议中，明文矩阵是使用压缩打包存储的，这需要在矩阵乘法之前进行扩展的预处理步骤，这非常耗时。 为了确保更公平的绩效评估，我们分析了多个投入的摊销成本。其中，ENSI是使用与NEXUS相同的SEAL库实现的，而ENSI*表示基于OpenFHE实现的结果。鉴于OpenFHE中的多线程并行性仍在开发中，所有实验都是在单核模式下进行的，以保持评估的一致性。如图\ref{time_pcmm}所示，不同输入尺度下明文密文矩阵乘法的性能比较表明：当输入尺度达到2^6$时，我们提出的方法的运行时间为1.41s，明显优于NEXUS 8.1\times$。随着投入规模扩大到2^11美元，性能优势仍然很大，比NEXUS快5.8美元。
\textbf{PCMM.} In NEXUS's plaintext-ciphertext matrix multiplication protocol, plaintext matrices are stored using compressed packing, necessitating an expanded pre-processing step before matrix multiplication, which is highly time-consuming. 
To ensure a fairer performance evaluation, we analyzed the amortized costs across multiple inputs. Among them, ENSI was implemented using the same SEAL library as NEXUS, while ENSI* represents results of implementation based on OpenFHE. Given that multi-threaded parallelism in OpenFHE is still under development, all experiments were conducted in single-core mode to maintain consistency across evaluations.
As illustrated in Figure \ref{time_pcmm}, the performance comparison of Plaintext-Ciphertext matrix multiplication under different input scales reveals that: when the input scale reaches $2^6$, our proposed method's runtime is 1.41s, significantly outperforming NEXUS by $8.1\times$. As the input scale expands to $2^{11}$, the performance advantage remains substantial, with $5.8\times$ faster than NEXUS.

\begin{figure}[!t]\centering
	\includegraphics[width=0.5\textwidth]{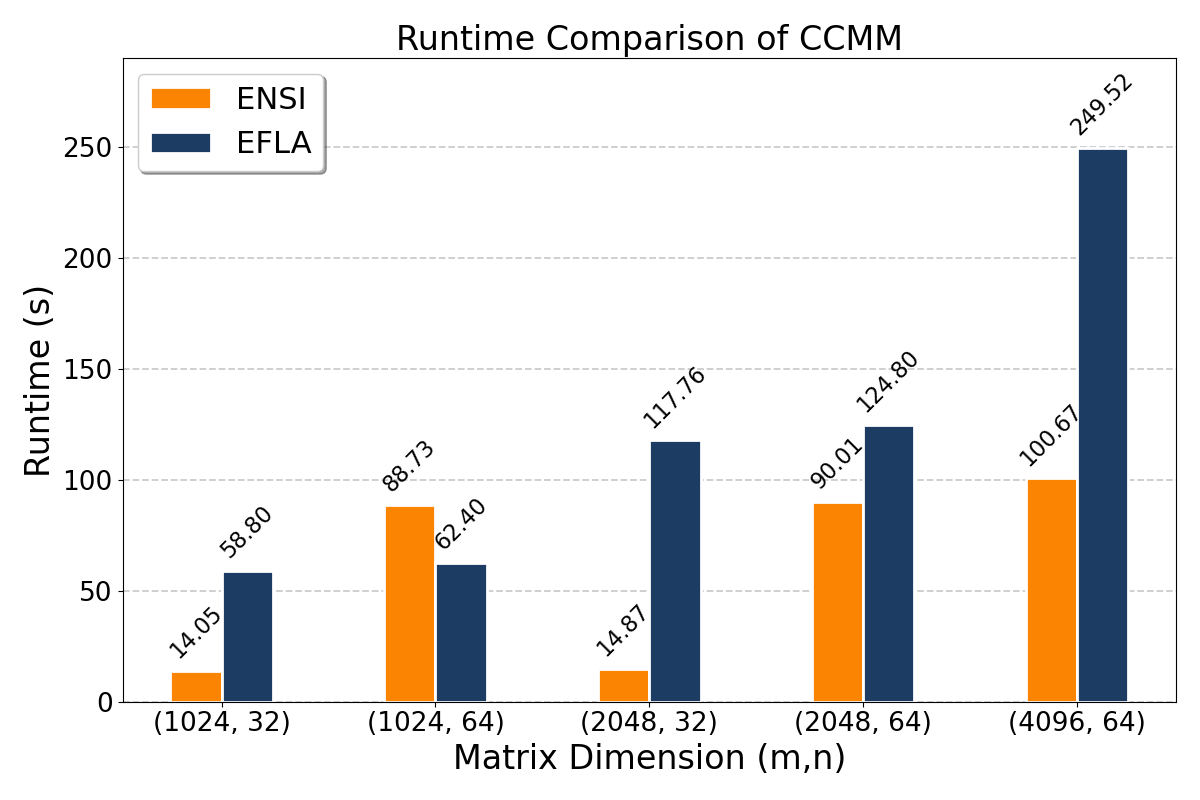}
	\caption{Evaluation of runtime for different ciphertext-plaintext matrix multiplication.}
    \label{time_ccmm}
\end{figure}

\textbf{CCMM.} As CCMM is not available in the open-source codebase of NEXUS, the performance evaluation was conducted by comparing ENSI and EFLA. Notably, the matrix multiplication in EFLA is implemented based on the JKLS algorithm\cite{jiang2018}. However, the JKLS algorithm requires the entire matrix to be packed within a single ciphertext, imposing stringent requirements on the scale of the parameter \(N^\prime\). Furthermore, in practical LLM deployment scenarios, matrix multiplication is typically performed by partitioning the matrices into smaller blocks and computing them separately.
Figure \ref{time_ccmm} presents the runtime overhead of ciphertext-ciphertext matrix multiplication (i.e., multiplication between matrices \(\mathbb{R}^{m\times n} \boxtimes \mathbb{R}^{n\times n}\)). The results indicate that as the matrix size increases, our proposed method demonstrates progressively greater computational efficiency. Additionally, due to the much higher computational overhead of CCMM compared to PCMM, EFLA is not included as a baseline in Figure \ref{time_pcmm}.

%Previous studies primarily focused on evaluating the performance of nonlinear functions GELU, LayerNorm, and Softmax in BERT models, while there has been limited research on the GELU variant GiLU and RMSNorm used in LLaMA models.
\textbf{Softmax.} Table~\ref{tab:soft} presents the secure performance evaluation of the Softmax function under our scheme, compared against the NEXUS method. Since Softmax converts score vectors into probability distributions, we conducted tests under varying vector dimensions and modulus $N^\prime$. We evaluate outputs in $\mathbb{R}^\ell$, with $\ell$ ranging from 1024 to 16384. The results show that when $N^\prime=14$, our method is approximately $2.2\times$ faster than NEXUS, and increasing to $N^\prime=16$, the performance improvement is more pronounced, ranging from $2.4\times$ to $2.6\times$.

% \begin{table}[ht]  
% \renewcommand{\arraystretch}{1.2}
%     \caption{Evaluation of Softmax function.}\label{tab:soft}
%     \centering  
%     % \begin{tabular}{l l r r r r r}  
%     \begin{tabular}{c c c c c c c}  
%         \hline  
%         \textbf{Setting} & \textbf{Protocol} & $1024$ & $2048$& $4096$ & $8192$ & $16384$ \\
%         \hline  
%         \multirow{4}{*}{$N=14$}   
%             & Ours       & 279  & 291   & 299  & 302 &$-$ \\
%             & NEXUS       & 635  & 661    & 675   & 689  &$-$\\
%             & Speedup       & $2.27\times$  & $2.26\times$  & $2.25\times$ & $2.25\times$ &$-$\\
%         \hline
%         \multirow{3}{*}{$N=16$}  
%             & Ours  & 1217   & 1220    & 1231   & 1243 & 1261  \\
%             & NEXUS      & 2954     & 2982    & 3162    & 3185 &3313\\
%              &Speedup    & $2.43\times$  & $2.44\times$  & $2.57\times$ & $2.56\times$ &$2.62\times$\\
%         \hline  
%     \end{tabular}  
% \end{table}  
\begin{table}[ht]  
\renewcommand{\arraystretch}{1.2}  
\caption{Evaluation of Softmax Function (\text{millisecond}).}  
\label{tab:soft}  
\centering  
\begin{tabular}{c c c c c c c}  
\toprule  
\textbf{Setting} & \textbf{Protocol} & $1024$ & $2048$& $4096$ & $8192$ & $16384$ \\
\midrule  
\multirow{3}{*}{$N^\prime=14$}  
    & NEXUS    & 635  & 661   & 675  & 689  & $-$ \\
    & ENSI     & 279  & 291   & 299  & 302  & $-$ \\
    & Speedup  & $2.27\times$  & $2.26\times$  & $2.25\times$ & $2.25\times$ & $-$ \\
\midrule  
\multirow{3}{*}{$N^\prime=16$}  
    & NEXUS    & 2954  & 2982  & 3162  & 3185 & 3313 \\
    & ENSI     & 1217  & 1220  & 1231  & 1243 & 1261  \\
    & Speedup  & $2.43\times$  & $2.44\times$  & $2.57\times$ & $2.56\times$ & $2.62\times$\\
\bottomrule  
\end{tabular}  
\end{table}  

\textbf{RMSNorm.} We evaluated the average approximation error under different parameter settings in single core environment following the RMSNorm experimental setup described in Table \ref{tab:rms_performance}. The average error is computed as the mean difference $\frac{\sum_{i=1}^{n} |y_i - y^\prime_i|}{n} $ between the true values $y_i \gets \Pi_\text{RMSNorm}(x_i), i\in[n]$ and their approximations $y^\prime_i \gets \Pi_\text{RMSNorm}(x_i)$. Since prior works have not addressed RMSNorm specifically, we do not include a runtime comparison. Experimental results demonstrate that, compared to NEXUS, our approximation scheme achieves comparable or superior accuracy. In NEXUS, the LayerNorm error is approximately 4.5e-4. Furthermore, we observe that although higher polynomial degrees increase circuit depth, they appear to better leverage fast optimization algorithms such as FFT and NTT, thereby reducing the overall computation time.

\begin{table}[htbp]  
\centering  
\renewcommand{\arraystretch}{1.2}  
\caption{Performance Comparison of Polynomial Approximation for RMSNorm Evaluation}  
\label{tab:rms_performance}  
\begin{tabular}{c| c c c c}  
\toprule  
\textbf{Input} & \textbf{Poly\_degree} & \textbf{Error} & \textbf{Runtime (s)} & \textbf{Level Usage} \\
\midrule  
\multirow{2}{*}{$\mathbb{R}^{2048\times128}$}   
    & 50 & 4.3e-5  & 6.8  & 17 \\
    & 60 & 4.0e-6  & 6.2  & 19 \\
\midrule  
\multirow{2}{*}{$\mathbb{R}^{4096\times512}$}   
    & 50 & 4.4e-5  & 26.1 & 17 \\
    & 60 & 3.5e-6  & 22.9 & 19 \\
\bottomrule  
\end{tabular}  
\end{table}

\subsection{End-to-End Benchmark.} 
Table \ref{tab:each-operations} presents the runtime breakdown for each individual operation within ENSI, based on the LLaMA-3-700M model with 16 layers. A total batch of 32 inputs was processed and benchmarked on a system equipped with a 32-core CPU and an A100 GPU. Runtime is the amortized latency per individual input. The inputs consist of commonly used sequences of 2048 tokens, representing the largest known scale of secure inference to date. In the table, values displayed in boldface correspond to runtimes measured on GPU. The “level” metric quantifies the hierarchical depth consumed by each operation. Our inference scale is three orders of magnitude larger than NEXUS, while achieving more than a twofold increase in inference speed. By integrating the bootstrapping operation into the RMSNorm, we have significantly reduced its computational overhead. In comparison, the bootstrapping operation accounts for $62.3\%$ of the total runtime in NEXUS, whereas in our approach it comprises only $1\%$.

\begin{table}[htbp]  
\centering  
\renewcommand{\arraystretch}{1.5}  
\vspace{0.5em}  
\caption{Performance Breakdown of Single-Token Generation Using ENSI.}  
\label{tab:each-operations}  
\resizebox{0.48\textwidth}{!}{  
\begin{tabular}{c| c| c |c}  
\toprule  
\textbf{Operation} & \textbf{Level}  & \textbf{Input} & \textbf{Times (s)} \\
\midrule  
\textsc{RMSNorm} & $19$ & $\mathbb{R}^{2048 \times 1536}$ & 3.01 \\
\textsc{Bootstrapping} & $20$ & $\mathbb{R}^{2048}$ & 0.04 \\
\textsc{PCMM} & $0$ & $(\mathbb{R}^{2048 \times 1536} \times  \mathbb{R}^{1536 \times 1536})\times 3$  & \textbf{1.78} \\
\textsc{RoPE} & $1$ & $\mathbb{R}^{2048 \times 96} \times 16$ & 1.61 \\
\textsc{CCMM} & $1$ & $ (\mathbb{R}^{2048 \times 96} \boxtimes  {\mathbb{R}^{2048 \times 96}}^{\mathrm{T}}) \times 16$ & \textbf{192.99} \\
\textsc{PCMM}$^\dagger$ & $1$ & $ (\mathbb{R}^{2048\times2048} \boxtimes  {\mathbb{R}^{2048}}) \times 16$ & \textbf{0.01} \\
\textsc{Softmax} & $5$ & $\mathbb{R}^{2048 \times 2048} \times 16$ & 93.61 \\
\textsc{CCMM} & $1$ & $(\mathbb{R}^{2048 \times 2048} \boxtimes  \mathbb{R}^{2048 \times 96}) \times 16$ & $\textbf{638.12}$ \\
\textsc{RMSNorm} & $19$ & $ \mathbb{R}^{2048 \times 1536} $ & 3.01 \\
\textsc{Bootstrapping} & $20$ & $\mathbb{R}^{2048}$ & 0.04 \\
\textsc{PCMM} & $0$ & $(\mathbb{R}^{2048 \times 1536} \boxtimes  \mathbb{R}^{1536 \times 4096}) \times 2$ & $\textbf{3.12}$ \\
\textsc{SiLU} & $6$ & $\mathbb{R}^{2048 \times 4096}$ & 14.11 \\
\textsc{CCMM}$^\dagger$ & $1$ & $(\mathbb{R}^{2048 \times 4096} \boxtimes  \mathbb{R}^{2048 \times 4096})$ & \textbf{0.01} \\
\textsc{RMSNorm} & $19$ & $\mathbb{R}^{2048 \times 4096}$ & 5.05 \\
\textsc{Bootstrapping} & $20$ & $\mathbb{R}^{2048}$ &  0.04 \\
\textsc{PCMM} & $0$ & $\mathbb{R}^{2048 \times 4096} \boxtimes  \mathbb{R}^{4096 \times 1536}$ &  $\textbf{1.57}$ \\
\bottomrule  
\end{tabular}  }
\vspace{0.5em} 
\par{$\dagger$ denotes the plaintext-ciphertext or ciphertext-ciphertext Hadamard product.}
\end{table}

\textbf{Accuracy}. Using the same hyperparameter settings as BitNet b1.58 \cite{Ma2024}, we evaluated ENSI’s zero-shot inference accuracy on the PIQA\cite{bisk2019}, COPA\cite{yeo2018copa}, and SST\cite{socher2013sst} datasets. These datasets collectively encompass a diverse range of tasks, including physical commonsense reasoning (PIQA), causal reasoning (COPA), and sentiment analysis (SST), thus providing a comprehensive assessment of the model’s performance and generalization across various reasoning scenarios. As shown in Table \ref{tab:model-performance}, ENSI’s inference accuracy is nearly comparable to that of plaintext inference.
We emphasize that the ideal objective of privacy-preserving inference is to match plaintext accuracy as closely as possible. Therefore, we do not include comparisons with other related methods and instead focus on the performance gap relative to the plaintext baseline.
%BoolQ\cite{clark2019boolq},
\begin{table}[ht]  
\centering  
\renewcommand{\arraystretch}{1.3}  
\caption{Inference accuracy performance on different datasets}  
\label{tab:model-performance}  
\begin{tabular}{c c c c}  
\toprule  
\textbf{Model} & \textbf{Dataset} & \textbf{Plaintext} & \textbf{ENSI} \\
\midrule  
\multirow{3}{*}{Bitnet b1.58-3B}  
    & PIQA  & $71.65\%$ & $65.89\%$ \\
    & COPA  & $72.12\%$ & $71.26\%$ \\
    % & BoolQ & $58\%$ & $51\%$ \\
    & SST   & $56.57\%$ & $55.08\%$ \\
    
\bottomrule  
\end{tabular}  
\end{table} 

\section{Conclusion}
% This paper proposes a novel non-interactive secure inference framework for LLMs. By synergistically integrating cryptographic schemes with large language model architectures, ENSI achieves efficient and secure inference.  ENSI introduces an optimized encoding strategy that effectively combines the CKKS with BitNet, along with improvements targeting matrix multiplication and non-linear function computations. Experimental results demonstrate that the proposed framework exhibits significant performance advantages. 

% Despite optimizations like matrix compression and batching, large scale matrix multiplication within the ciphertext domain remains the dominant  bottleneck. Currently, secure inference for LLMs still equires substantial computational resources and exhibits high latency. In the future, we aim to deeply integrate dedicated hardware acceleration with the ENSI framework, achieving fully secure large model inference on GPUs.
This paper presents ENSI, a non-interactive, privacy-preserving inference framework for large language models that synergistically integrates advanced cryptographic schemes with LLM architectures. ENSI introduces an optimized encoding scheme that effectively unifies CKKS homomorphic encryption with BitNet while simultaneously refining ciphertext-level matrix multiplications and encrypted non-linear activations.
Our experimental results demonstrate that ENSI outperforms existing solutions and exhibits significant performance advantages.
In addition, the adaptation to BitNet occurs primarily during the plaintext training phase. Since our secure inference protocol is applied post-training, the additional effort required for migration is relatively minimal.

However, even with optimizations like matrix compression and batching, large-scale ciphertext matrix multiplication remains the primary bottleneck. Secure LLM inference still requires significant computational resources and incurs high latency. Looking forward, we aim to deeply integrate dedicated hardware acceleration with the ENSI framework, achieving fully secure large model inference on GPUs.

% However, despite optimizations such as matrix compression and batching, large-scale matrix multiplication within the ciphertext domain remains the dominant  bottleneck. Currently, secure inference for LLMs still equires substantial computational resources and exhibits high latency.  Looking forward, we aim to deeply integrate dedicated hardware acceleration with the ENSI framework, achieving fully secure large model inference on GPUs.

\section*{Acknowledgment}
This work is supported by National Nature Science Foundation of China (No.62472431) and Natural Science Foundation of Hunan Province, China (GrantNo.2023JJ30640).

\normalem
\bibliographystyle{IEEEtran}
\bibliography{main}

% \vspace{12pt}
% \color{red}
% IEEE conference templates contain guidance text for composing and formatting conference papers. Please ensure that all template text is removed from your conference paper prior to submission to the conference. Failure to remove the template text from your paper may result in your paper not being published.

\end{document}